\documentclass[preprint,12pt]{elsarticle}

\usepackage{amssymb}
\usepackage{amsthm}
\usepackage{amsmath,bm}
\usepackage[utf8]{inputenc}
\usepackage[T1]{fontenc}
\usepackage[section]{placeins}
\usepackage{float}
\usepackage{graphicx}
\usepackage{subcaption}
\usepackage{hyperref}
\hypersetup{
    colorlinks=true,
    linkcolor=blue,
    filecolor=magenta,      
    urlcolor=cyan,
    pdftitle={Overleaf Example},
    pdfpagemode=FullScreen,
    }

\journal{European Journal of Mechanics -B/Fluids}

\begin{document}

\begin{frontmatter}



\title{Symmetry-reduced low-dimensional representation of large-scale dynamics in the asymptotic suction boundary layer}

\author[inst1]{Matthias Engel
\corref{cor1}}
\cortext[cor1]{Corresponding authors}
\author[inst2]{Omid Ashtari}
\author[inst1]{Moritz Linkmann
\corref{cor1}}

\affiliation[inst1]{organization={School of Mathematics and Maxwell Institute for Mathematical Science, University of Edinburgh},
            addressline={Mayfield Rd}, 
            city={Edinburgh},
            postcode={EH9 3FD}, 
            country={United Kingdom}}

\affiliation[inst2]{organization={Emergent Complexity in Physical Systems Laboratory, \'{E}cole Polytechnique F\'{e}d\'{e}rale de Lausanne},Department and Organization
            city={Lausanne},
            postcode={CH-1015}, 
            country={Switzerland}}

\begin{abstract}


An important feature of turbulent boundary layers are persistent large-scale coherent structures in the flow.  Here, we use Dynamic Mode Decomposition (DMD), a data-driven technique designed to detect spatio-temporal coherence, to
construct optimal low-dimensional representations of such large-scale dynamics
in the asymptotic suction boundary layer (ASBL).  In the ASBL, fluid is removed
by suction through the bottom wall, resulting in a constant boundary layer
thickness in streamwise direction.  
That is, the streamwise advection of coherent
structures by the mean flow ceases to be of dynamical importance and can be
interpreted as a continuous shift symmetry in streamwise direction.  However,
this results in technical difficulties, as DMD is known to perform poorly in
presence of continuous symmetries. We address this issue using symmetry-reduced DMD (Marensi {\em et al.}, J. Fluid Mech. {\bf 721}, A10 (2023)), and find the large-scale dynamics of the ASBL to be low-dimensional indeed and potentially self-sustained, featuring ejection and sweeping events at large scale. Interactions with 
near-wall structures are captured when including only a few more modes.

\end{abstract}




\end{frontmatter}


\section{Introduction}


Turbulent superstructures, that is, large-scale spatio-temporally coherent structures, occur at high Reynolds number in a variety of turbulent flows. Some of the prime examples were found experimentally in turbulent boundary layers  
\cite{Meinhart,Kim1999,Adrian2000,Guala2006,Hutchins2007a}. They influence momentum-transport and mixing processes~\cite{Robinson1991}, 
and carry most of the kinetic energy of the flow \cite{Monty2007,Hutchins2011} and the Reynolds shear stress \cite{Ganapathisubramani2003,Guala2006}. Even though turbulent superstructures occur at considerable distances from the wall, they interact and affect the dynamics of small-scale, wall-attached structures \cite{Hutchins2007b,Bross2019}. 
How exactly this interaction proceeds, and how and if the small scales directly influence the large scales is an open question. Similarly, the mechanism maintaining large-scale coherent structures is currently not known.
Here, we consider these questions at moderate Reynolds number in a broader sense of spatio-temporal coherence at or close to system scale. As an example flow we consider the asymptotic suction boundary layer (ASBL), where the boundary layer thickness remains constant in streamwise direction as a result of suction applied to the bottom wall,
and discuss and apply a method for extraction and description of thereof.




A flexible way to extract dynamically interesting flow features is dynamic mode decomposition (DMD)\cite{schmid_2008,schmid_2010}. It is a data-driven method related to Koopman analysis for nonlinear dynamical systems~\cite{Mezi2005SpectralPO, mezic2013}. DMD can be applied to a time series of snapshot pairs taken from flow fields obtained by direct numerical simulation (DNS) or experiment and results in a best-fit, usually high-dimensional, linear operator $\bm{A}$ that maps one data snapshot to the next separated by a fixed time interval. The spectral analysis of the linear operator $\bm{A}$ provides us with a decomposition of the nonlinear dynamical system in basis functions, or dynamic modes, which evolve exponentially in time and contain spatial structures of the flow.  

The advantages of DMD come with a drawback. Since it is based on singular value decomposition (SVD), its performance deteriorates in presence of continuous, e.g. shift, symmetries in data~\cite{Kutz2016}. Convergence issues may occur and more generally, the linear operator $\bm{A}$ will include features of the symmetries that are dynamically irrelevant. This leads to spurious dynamic modes and   
hence adversely affects the reconstruction and detection of physically important flow features. 
Several approaches have been developed to address this problem. 
In physics-informed DMD the analysis is performed on a fast Fourier transformation of the translation invariant flow fields to decouple the wave numbers responsible for drifts in streamwise or spanwise directions~\cite{pidmd}. For time-independent drifts, characteristic DMD (CDMD) extracts the dynamic modes in a properly chosen frame of reference~\cite{characteritic_dmd}. Recently, Marensi {\em et al.} combined a method for symmetry reduction that had been very useful in the dynamical systems approach to turbulence, the {\em method of slices} with DMD. In the method of slices a state-space trajectory corresponding to the fluid flow is projected onto a symmetry-reduced state space, in which every group orbit consisting of symmetry-equivalent points is represented by only one representative point, that is, one operates in the factor space on a chosen representative of an equivalence class~\cite{SIMINOS2011187, PhysRevLett.114.084102}. Symmetry reduction using the method of slices has already carried out successfully for high-dimensional dynamical systems, such as numerical simulations of pipe flow ~\cite{willis_cvitanović_avila_2013, PhysRevE.93.022204} and recently in combination with DMD to plane Couette and plane Poiseuille flow simulations up to friction Reynolds number $Re_\tau = 205$ with translation symmetries in streamwise and spanwise direction \cite{marensi_yalniz_hof_budanur_2023}.

In this paper, we apply symmetry-reduced DMD (SRDMD \cite{marensi_yalniz_hof_budanur_2023}) to data obtained from DNS of the ASBL at  $Re_\tau\approx 330$, in order to construct low-dimensional representations of the dynamics of a large-scale persistent coherent structure in the flow. This large-scale low-momentum region extends through the simulation domain in streamwise direction and drifts transversely through the computational domain. The aim is to describe and understand the large-scale dynamics of the ASBL and its interaction with the small scales. We find that a four-dimensional representation is sufficient to describe the main dynamical features of the coherent structure. Therein, we find sweep and ejection events.  Interestingly, this does not involve near-wall small-scale structures. A seven-dimensional representation captures the interaction between the large-scale coherent structure and small-scale wall-attached structures. 

The paper is structured as follows. In sec.~\ref{sec:methods}, we provide a summary of the first Fourier mode slicing technique used for the reduction of the continuous symmetries that underlie our investigated flow systems. Subsequently, we discuss sparsity-promoting DMD with its advantages over the classical DMD approach when selecting proper modes for the reconstruction of the original time series. Section \ref{sec:asbl} introduces the ASBL.
In sec.~\ref{sec:3drotasbl} we illustrate symmetry-reduced sparsity-promoting DMD on a test case consisting of rotating ASBL system at a low Reynolds number, 
with a continuous symmetry in spanwise direction leading to a slow but time-dependent shift. 
Section \ref{sec:2dasbl} contains our analysis of the ASBL at 
$Re_\tau \approx 330$ with time-dependent continuous symmetries in streamwise and spanwise direction leading to a transverse drift through the computational domain. We conclude with a summary of our results and provide suggestions for further work in sec.~\ref{sec:conclusions}.

\section{Methods}
\label{sec:methods}
\subsection{Symmetry reduction - first Fourier mode slice approach}\label{sec:slicing}
Having in mind its application to the ASBL, we introduce the slicing method in channel geometry following Ref.~\cite{marensi_yalniz_hof_budanur_2023}. That is, we consider flow fields periodically extended in streamwise direction, $x$, and spanwise direction, $z$, and bounded in the wall-normal direction, $y$. We numerically represent a flow field in the form of a Fourier $\times$ Chebyshev $\times$ Fourier basis expansion:

\begin{equation}
    \bm{u}(\bm{x})=\sum_{k_x}\sum_{n}\sum_{k_z}\bm{\tilde{u}}_{k_x,n,k_z}T_n(y)e^{2\pi i(k_xx/L_x+k_zz/L_z)},
\end{equation}
where $\bm{\tilde{u}}_{k_x,n,k_z}$ are complex Fourier/Chebyshev coefficients and $T_n$ the $n^{\text{th}}$ Chebyshev polynomial as a function of the coordinate in wall-normal direction. The domain sizes in streamwise and spanwise direction are denoted by $L_x$ and $L_z$, respectively. The flows we investigate here have continuous symmetries in streamwise and spanwise direction leading to dynamically irrelevant degrees of freedom causing translations in streamwise and spanwise direction. As mentioned in the introduction, the resulting equivalent copies of the flow states contained in the dataset cause the data approximation obtained with the SVD-based DMD, and with that the quality of any low-dimensional representation, to deteriorate. Therefore, it is necessary to align the data before. Budanur {\em et al.}~\cite{PhysRevLett.114.084102} proposed a method for the symmetry reduction of systems with Fourier space discretisation and SO(2) symmetry using the relation of a flow field $\bm{u}(\bm{x})$ shifted by $\delta \bm{x}$ with the rotation of the state space vector containing the flow field's Fourier coefficients. The symmetry reduction is then obtained by a phase-fixing transformation that, in the presence of two continuous translation symmetries, restricts the state space trajectory on a co-dimension 2 submanifold. In practise, this transformation is computed locally on a linear approximation to the submanifold called a {\em slice}, for which a proper coordinate system has to be found. 

The slicing method has recently been extended to channel geometry, allowing to reduce symmetries in stream- and spanwise directions \cite{marensi_yalniz_hof_budanur_2023}. The slice coordinate system is constructed using two  template functions
\begin{equation}\label{equ:template}
    \bm{\hat{u}_x'}:=f(y)\cos(2\pi x/L_x)\bm{\hat{x}} \quad \text{and} ~\quad\bm{\hat{u}_z'}:=f(y)\cos(2\pi z/L_z)\bm{\hat{x}},
\end{equation}
in which a template profile $f(y)$ as a function of the wall-normal direction has to be found. This choice of template functions and their spatial gradients define a coordinate system in the state space, in which the shift of a flow field along one of its homogeneous directions is represented by a rotation of its state space vector by an angle $\hat\phi$. Therefore, the transformations that reduce the translation symmetries can be defined as
\begin{equation}\label{equ:symredtrafo}
    S_x(\bm{u}):=g(-(\hat{\phi}_x/2\pi)L_x,0)\bm{u} ~~ \text{and} ~~ S_z(\bm{u}):=g(0,-(\hat{\phi}_z/2\pi)L_z)\bm{u},
\end{equation}
which fix the polar angles 
\begin{equation}\label{equ:polaranglex}
    \hat{\phi}_x:=\arg(\left\langle\bm{u},\bm{\hat{u}'}_x\right\rangle+i\left\langle\bm{u},g(L_x/4,0)\bm{\hat{u}_x'}\right\rangle)
\end{equation}
and
\begin{equation}\label{equ:polaranglez}
    \hat{\phi}_z:=\arg(\left\langle\bm{u},\bm{\hat{u}'}_z\right\rangle+i\left\langle\bm{u},g(0,L_z/4)\bm{\hat{u}_z'}\right\rangle)\ ,
\end{equation}
where the angled brackets denote the $L_2$-inner product on the domain. 
The operator $g(\delta x,\delta z)$ defines the translation in $x$- and $z$-direction by $\delta x$ and $\delta z$, respectively. 

In order to obtain well-defined polar angles (\ref{equ:polaranglex}-\ref{equ:polaranglez}) at all times, we have to prevent their corresponding complex number to vanish. For that it is crucial to find the optimal template profile $f(y)$ and the direction the template fields should be aligned to. An important property to consider here are the discrete symmetries of the involved flow fields, which can lead both arguments of the polar angles to vanish identically. That is the symmetries of the template fields in eq.~(\ref{equ:template}) determined by the template profile $f(y)$, the symmetries of the cosine or sine function, respectively, and any discrete symmetry of the flow fields must be taken into account. 

Alternatively, Rowley {\em et al.}~\cite{ROWLEY20001} showed that the changes of the shifts $\delta x$ and $\delta z$ in time follow from an optimisation, called template fitting, that matches the data with the chosen template functions. Using the relations $\phi_x(t)=2\pi\delta x(t)/{L_x}$ and $\phi_z(t)=2\pi\delta z(t)/{L_z}$ the phases $\hat\phi_x$ and $\hat\phi_z$ can be obtained by integrating the reconstruction equations

\begin{equation}\label{equ:reconstructionequx}
    \dot{\hat{\phi}}_x(t)=\left(\frac{2\pi}{L_x}\right)\frac{\left\langle\partial_x \bm{\hat{u}}^{'}_x, \partial_t\bm{u}|_{\bm{u}=\bm{\hat{u}}}(t)\right\rangle}{\left\langle\partial_x\bm{\hat{u}}^{'}_x, \partial_x\bm{\hat{u}}(t)\right\rangle}
\end{equation}

\begin{equation}\label{equ:reconstructionequz}
    \dot{\hat{\phi}}_z(t)=\left(\frac{2\pi}{L_z}\right)\frac{\left\langle\partial_z \bm{\hat{u}}^{'}_z, \partial_t\bm{u}|_{\bm{u}=\bm{\hat{u}}}(t)\right\rangle}{\left\langle\partial_z\bm{\hat{u}}^{'}_z, \partial_z\bm{\hat{u}}(t)\right\rangle}.
\end{equation}
Additionally, we can use the behaviour of the reconstruction equations (\ref{equ:reconstructionequx}-\ref{equ:reconstructionequz}) to identify cases in which the trajectory reaches the slice border causing the reconstruction equations to become singular and the slicing technique to fail. These cases lead to nonphysical jumps of the flow fields in their shift directions.


\subsection{Sparsity-promoting Dynamic Mode Decomposition}

The numerical data we process consists of an equidistant time series of $N$ spatially resolved flow fields $\bm{u}(t_j)$, called snapshots, at discrete times $t_j$ for $j\in\{1,2,...,N\}$. Denoting with $f(\bm{u}(t_j))$ the $j^{\text{th}}$ state space column vector that represents the flow field $\bm{u}(t_j)$, we construct a data matrix 

\begin{equation}
    \bm{X}=[f(\bm{u}(t_{i_1})),..,f(\bm{u}(t_{i_{M}}))],
\end{equation}
consisting of a subset of $M$ snapshots, labelled with $i$, which can be drawn at random in time. Additionally, we set up a second data matrix

\begin{equation}
    \bm{X'}=[f(\bm{u}(t_{i_1}+\delta t)),...,f(\bm{u}(t_{i_M}+\delta t))],
\end{equation}
consisting of the state vectors from the data matrix $X$, but shifted in time by a fixed $\delta t$. We are now interested in the best fit linear operator $\bm{A}$ that maps $\bm{X}$ to $\bm{X}'$, which is given by

\begin{equation}
    \bm{A}=\bm{X'}\bm{X}^{+},
\end{equation}
where $\bm{X}^{+}$ denotes the Moore-Penrose pseudo inverse of $\bm{X}$, obtained by applying singular value decomposition (SVD) of the data matrix $\bm{X}=\bm{U}\bm{\Sigma}\bm{V}^{*}$ leading to

\begin{equation}\label{equ:svd}
    \bm{A}=\bm{X}'\bm{V}\bm{\Sigma}^{-1}\bm{U}^{*},
\end{equation}
where $^{*}$ indicates Hermitian transposition. The eigendecomposition of $\bm{A}$ provides us with dynamic modes and frequencies that we can use to identify dominant oscillatory structures of the flow, even if recurrent flow events have not yet taken place. When building a lower-dimensional representation of the flow, we approximate the flow fields by superimposing $m \leqslant M$ of the $M$ calculated dynamic modes, using their corresponding DMD frequencies as well as the amplitudes of the dynamic modes. Taking complex conjugates into account, we obtain   

\begin{equation}\label{equ:dmdapprox}
    \bm{u}(t_j)\approx\sum_{k=m-1}^{m-1}a_{i_k}e^{(\sigma_{i_k}+i\omega_{i_k})t_j}\bm{\psi}_{i_k}.
\end{equation}
Here, $\bm{\psi}_{i_k}$ is the $i_k^{\text{th}}$ dynamic mode, $\sigma_{i_k}\in\mathbb{R}$ its temporal growth or decay rate, $\omega_{i_k}\in\mathbb{R}$ its oscillation frequency, and $a_{i_k}\in\mathbb{C}$ its amplitude. The latter is not returned by the DMD analysis itself and still has to be determined. Furthermore, it is not {\em a priori} clear, which subset of dynamic modes is optimal to represent the data set. A common approach is to use an SVD-truncated version of (\ref{equ:svd}) and to choose the most energetic dynamic modes for the reconstruction. Despite the prevention of over-fitting and therefore the occurrence of nonphysical dynamic modes, it raises the additional question of what the optimal truncation number is and it does not necessarily provide the subset of dynamic modes that gives the best approximation result. This impasse is overcome by using sparsity-promoting DMD \cite{Jovanovic2012SparsitypromotingDM}, which combines a convex optimisation to determine the optimal vector of DMD amplitudes $\bm{a}\in\mathbb{C}^{2M-1}$ and a penalisation of the number of its non-zero components, 

\begin{equation}\label{equ:spDMD}
    \min_{\bm{a}} J(\bm{a}) + \gamma\sum_{k=-M+1}^{M-1}|a_k|,
\end{equation}
where $J(\bm{a})$, is the squared average $L_2$-distance of the DMD approximation from the original data, 

\begin{equation}\label{equ:minimisel2distance}
    J(\bm{a}):=\frac{1}{M}\sum_{m=0}^{M-1}\big\| \bm{u}(\bm{x},t_{i_m}))-\sum_{k=-M+1}^{M-1}a_ke^{(\sigma_k+i\omega_k)t_{i_m}}\bm{\psi}_k\big\|^2,
\end{equation}
and $\gamma > 0$ is the sparsity controlling regularisation parameter. The value of $\gamma$ determines the necessary compromise between the number of non-zero elements of the amplitude vector $\bm{a}\in\mathbb{C}^{2M-1}$ and the quality of the least-squares approximation for $J(\bm{a})$ in (\ref{equ:spDMD}). 

In case the underlying data shadows the trajectory of an oscillatory structure the sparsity-promoting DMD provides $n+1\leqslant M$ non-zero DMD amplitudes for the neutral mode and the modes related to the fundamental frequency and its $n-1$ harmonics. After the reconstruction, we measure the quality of the sparsity-promoting DMD approximation for an oscillatory structure by estimating a fundamental frequency for our problem as
\begin{equation}\label{equ:Tg}
    \omega_f(n):=\frac{2}{n(n+1)}\sum_{m\in\{m_1,..,m_n\}} \omega_m,
\end{equation}
with its residual defined as \cite{page_kerswell_2020}
\begin{equation}\label{equ:residualTg}
    \epsilon_\omega(n):=\frac{1}{n|\omega_f|^2}\sum_{m\in\{m_1,..,m_n\}}|\omega_m-m\omega_f|^2.
\end{equation}

\subsection{Implementations}

For all numerical simulations presented here, we used the open source library {\tt channelflow2.0}, which was developed originally by John F. Gibson~\cite{channelflow} to solve the Navier-Stokes equations for incompressible flows in channel geometry with no-slip boundary conditions at the bottom and top, and periodic boundary conditions in stream- and spanwise directions. We translated the first Fourier mode slicing technique as well as the Matlab implementation of sparsity-promoting DMD based on~\cite{Jovanovic2012SparsitypromotingDM} into C\texttt{++} in order to use the {\tt channelflow2.0}~\cite{channelflow} infrastructure.

\section{Asymptotic suction boundary layer}
\label{sec:asbl}
The ASBL is an open, incompressible flow over a porous plate through which the fluid with the free-stream velocity $U_{\infty}$ is sucked at a constant suction velocity $V_S$ perpendicular to the plate. Far from the leading edge of the plate, this suction prevents the boundary layer to grow and causes the Reynolds number being independent from the location on the plate.
We refer to a Cartesian coordinate system in which we designate the downstream, wall-normal and spanwise directions with $x$, $y$ and $z$. The equations of motion for the incompressible flow are the Navier-Stokes equations
\begin{equation}
\label{eq:momentum}
    \partial_t\bm{u}+(\bm{u}\cdot\nabla)\bm{u}=-\nabla p+\nu\Delta\bm{u},
\end{equation}
together with the continuity equation
\begin{equation}
    \nabla\cdot\bm{u}=0, 
\end{equation}
where $\bm{u}$, $p$ and $\nu$ are velocity, pressure divided by the constant density, and kinematic viscosity, respectively.
Imposing the boundary conditions
\begin{equation}
    \bm{u}(x,y=0,z)=(0,-V_S,0), ~ \bm{u}(x,y={\infty},z)=(U_{\infty},-V_S,0),
\end{equation}
we obtain the laminar velocity profile for the ASBL,
\begin{equation}\label{equ:asbl-profile}
    \bm{u}_0=\left(U_{\infty}(1-e^{-y/\delta}),-V_S,0\right),
\end{equation}
determined from the equations of motion. For this half-open flow, the laminar displacement thickness $\delta=\nu/V_S$ is a length scale that can be used to define the Reynolds number. If we use the free-stream velocity $U_{\infty}$ as the characteristic velocity scale, we obtain
\begin{equation}
    Re=\frac{U_{\infty}\delta}{\nu}=\frac{U_{\infty}}{V_S}.
\end{equation}

In what follows, velocities are given in units of $U_{\infty}$ and the time in units of $\delta/U_{\infty}$ and all lengths are measured in units of $\delta$. For numerical treatment the open flow is restricted to a finite computational domain by using a second plate at a distance $h$ from the lower plate, which moves at a constant velocity $U_{\infty}$ in positive direction $x$, emulating the free-stream velocity of the open flow, while the bottom plate rests. In order to ensure the conservation of mass, the fluid must be introduced into the system through the upper plate in the same way as it is sucked through the lower one. In spanwise and streamwise directions, we assume periodic boundary conditions with periods $L_x$ and $L_z$, respectively. Thus, computational domain is rectangular of dimension $L_x\times h\times L_z$. Compared to equation (\ref{equ:asbl-profile}) the velocity profile for the laminar flow in streamwise direction changes due to the new boundary condition at the upper plate. We obtain now
\begin{equation}
     \bm{u}_0=\left(U^{*}_{\infty}(1-e^{-y/\delta}),-V_S,0\right),
\end{equation}
where $U^*_{\infty}=U_{\infty}/(1-e^{-h/\delta})$. 

\section{Method validation and test case}\label{sec:3drotasbl}

In dealing with symmetry reductions the first Fourier mode slicing technique becomes especially helpful when the drift caused by the underlying continuous symmetries is time dependent. In order to verify the implemented methods we constructed a simple test case that carries interesting flow features for our analysis that are similar to the flow features of the ASBL that we are mainly interested in. To achieve that, we considered a spanwise rotating ASBL at a low Reynolds number $\text{Re}=400$ in a small periodic domain of $L_x/\delta\times h/\delta\times L_z/\delta=2\pi\times 5 \times 0.77\pi$. Rotation was implemented by addition of the Coriolis term $2 \bm{u} \times \Omega \bm{e}_z$, where $\bm{e}_z$ is the unit vector in spanwise direction and $\Omega$ the rotation rate, on the right-hand side of the momentum equation \eqref{eq:momentum}. The pressure now includes the centrifugal force $-|\Omega \bm{e}_z \times \bm{x}|^2$. 

\begin{figure}[t]
\begin{minipage}[b]{0.6\textwidth}
\begin{subfigure}{\textwidth}
\subcaption{}
\includegraphics[width=\linewidth]{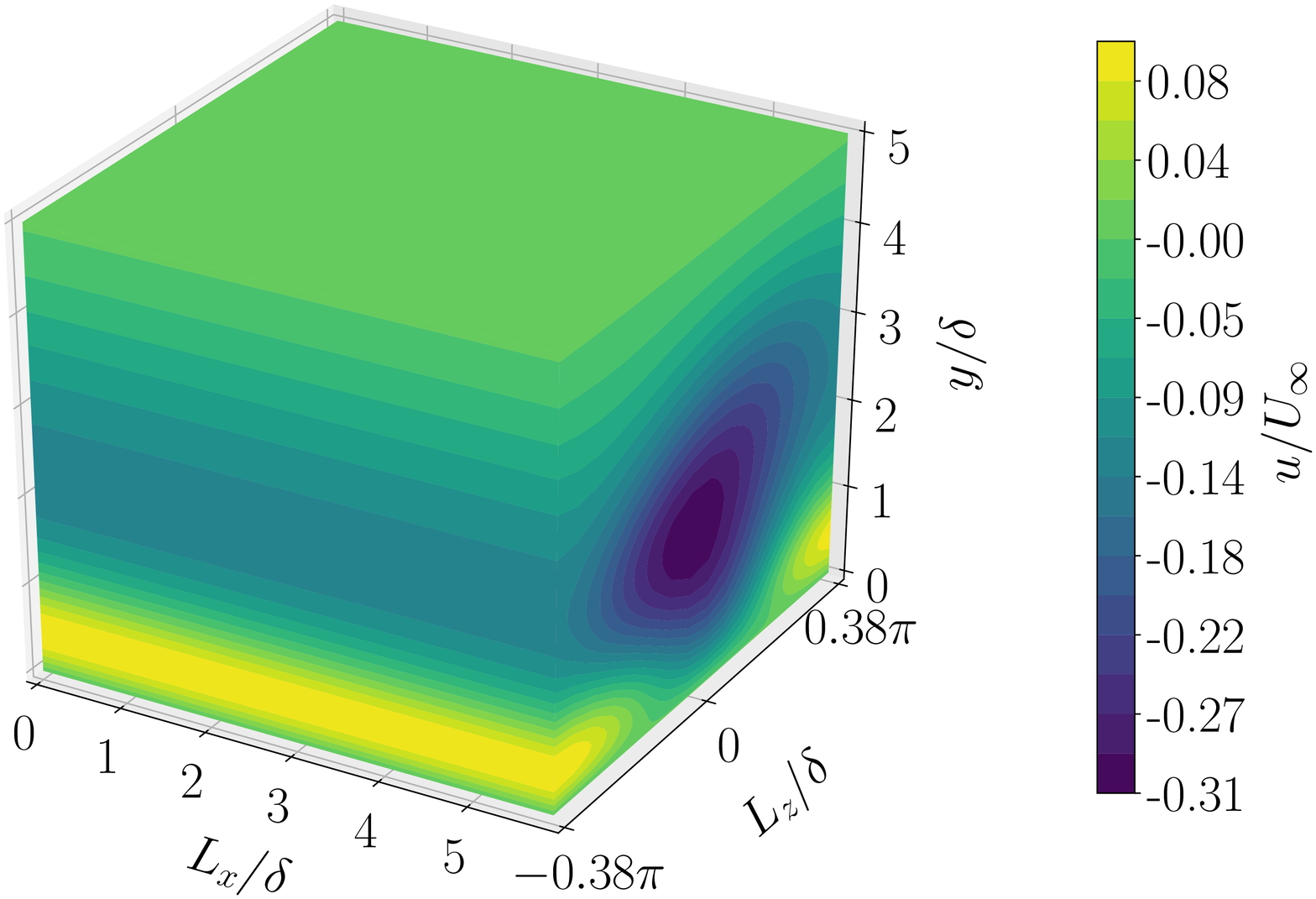}
\end{subfigure}\par
\end{minipage}
\begin{minipage}[b]{0.4\textwidth}
\begin{subfigure}{\textwidth}
\subcaption{}
\includegraphics[width=\linewidth]{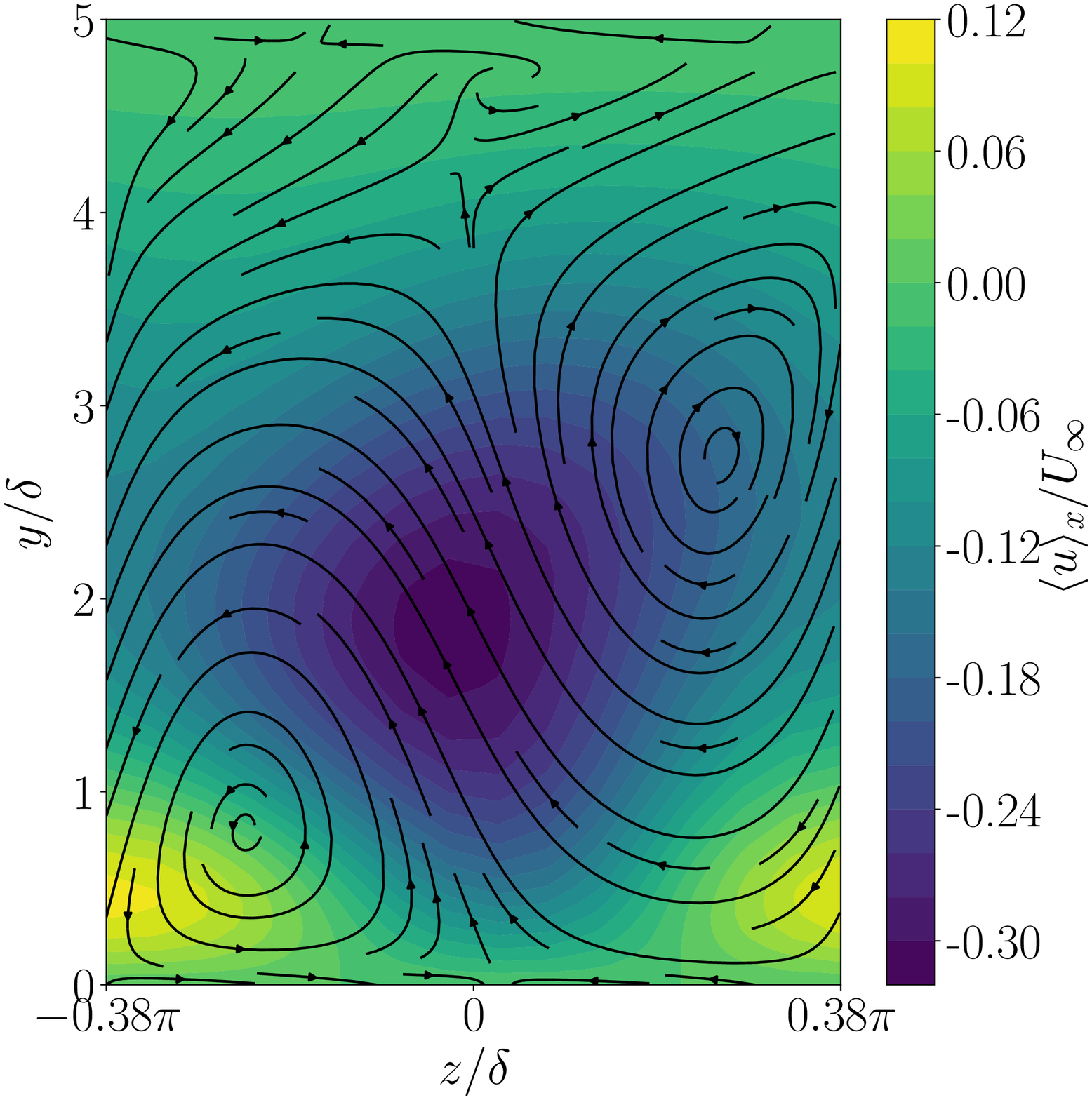}
\end{subfigure}
\end{minipage}
\quad
\begin{minipage}[b]{\textwidth}
\caption{Streamwise velocity of a snapshot taken from the time series as deviation from the laminar profile shown as (a) 3d contour plot and (b) averaged in streamwise direction together with the $y$-$z$ streamfunction. The colour code indicates a low-momentum region in blue extending into the free stream and two near-wall high-momentum regions in yellow.}\label{fig:3dbox}
\end{minipage}
\end{figure}

At a rotation rate $\Omega=0.003U_{\infty}/\delta$ the laminar state becomes unstable, leading to the formation of a stable relative periodic orbit (SRPO) dominated by a free-stream low-momentum region with its center at wall-normal height of about $y/\delta\approx 1.9$ and two near-wall high-speed streaks located to the left an right of the low-momentum region near the bottom wall, shown in Figure \ref{fig:3dbox}(b). To ensure the flow settled down, we followed the trajectory for 10,000 advective time units. The system has a continuous symmetry in spanwise direction leading the low-momentum region to drift in spanwise direction. 

\begin{figure}[t]
\begin{minipage}[b]{0.5\textwidth}
\begin{subfigure}{\textwidth}
\subcaption{}
\includegraphics[width=\linewidth]{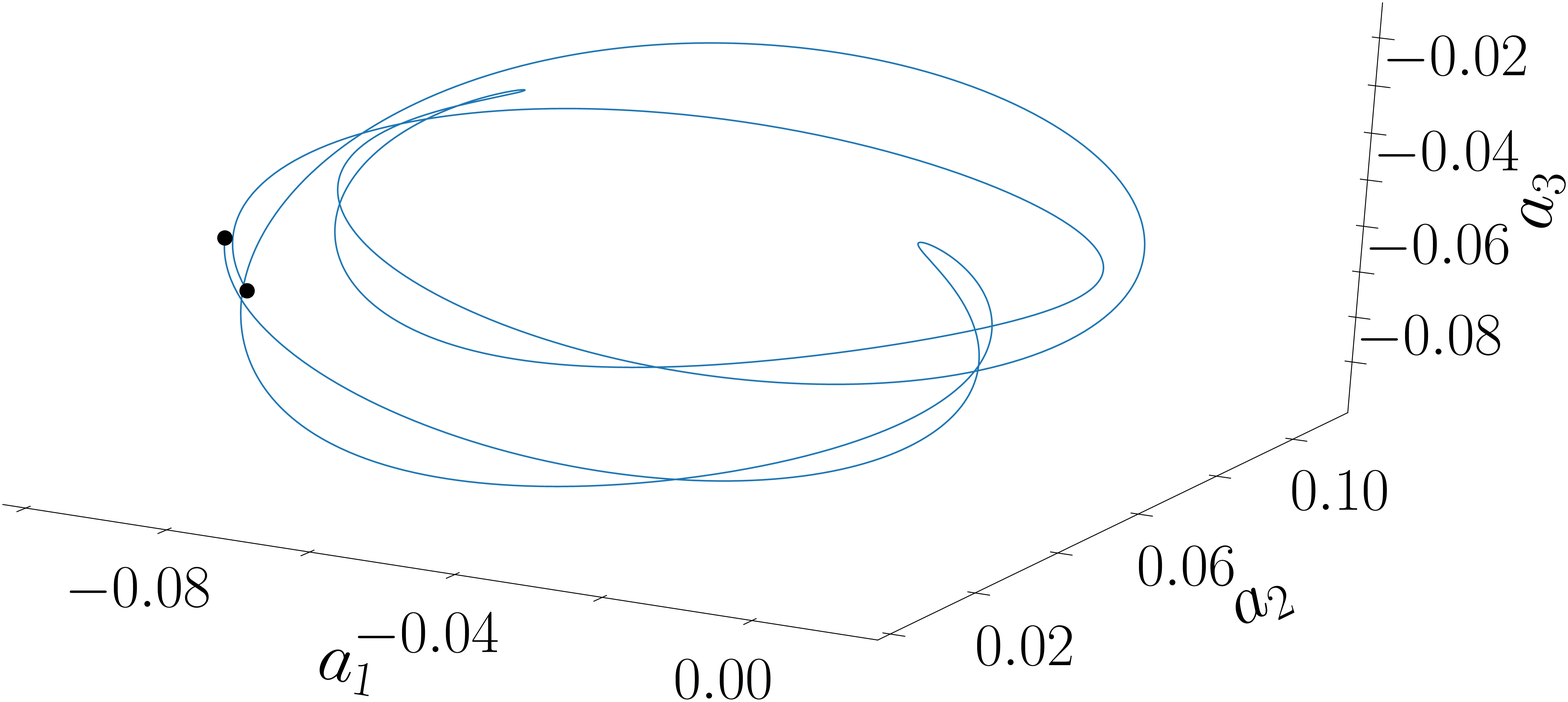}
\end{subfigure}\par
\end{minipage}
\begin{minipage}[b]{0.5\textwidth}
\begin{subfigure}{\textwidth}
\subcaption{}
\includegraphics[width=\linewidth]{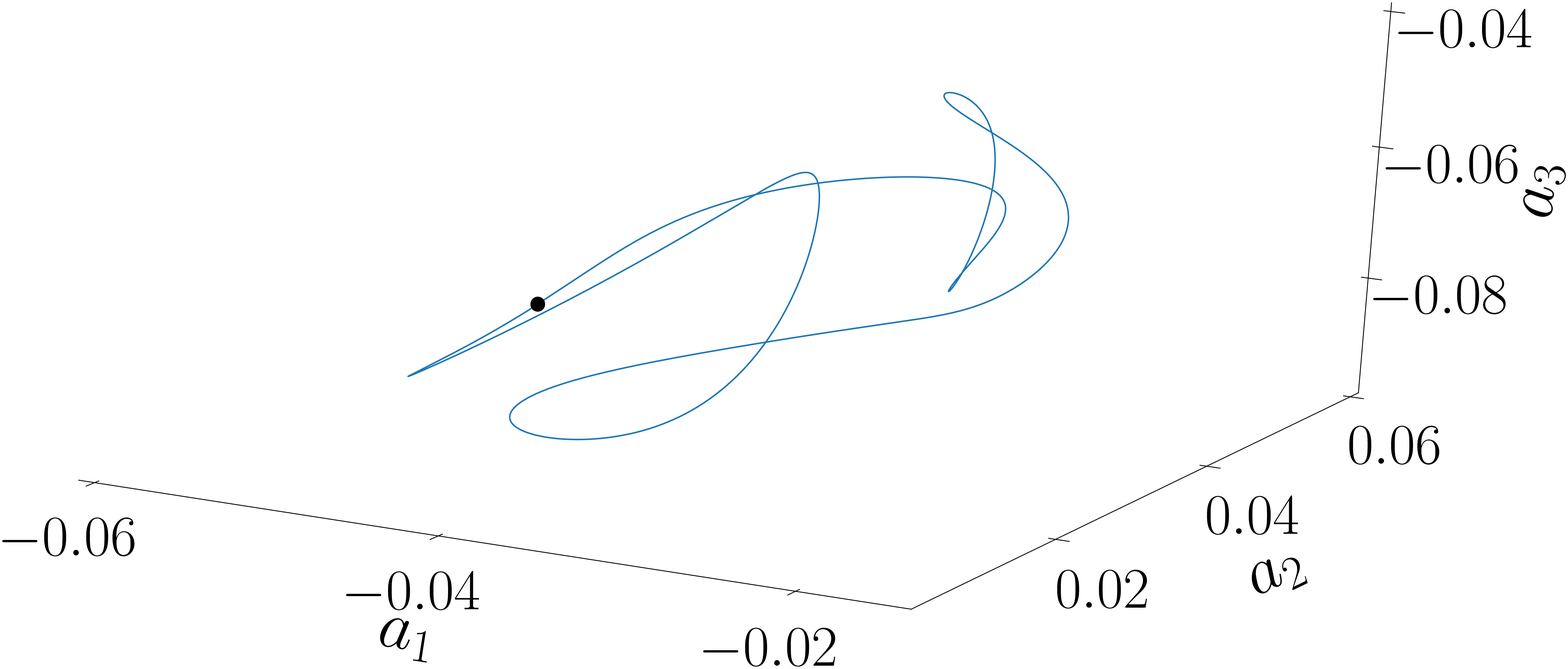}
\end{subfigure}
\end{minipage}
\quad
\begin{minipage}[b]{\textwidth}
\caption{State space projections for the trajectory of the SRPO over about one period. (a) represents the trajectory of the SRPO contained in the original data set and (b) the symmetry-reduced trajectory after applying the first Fourier mode slicing technique. The black dots indicate start and end points of the trajectories. The projection $a_i=\left\langle\bm{u},\bm{e}_i\right\rangle$ was performed onto orthogonal unit vectors $\bm{e}_i$ created from linearly independent flow fields of the DNS.}\label{fig:trajectories}
\end{minipage}
\end{figure}

The phase portraits shown in Figure~\ref{fig:trajectories} illustrate the effect of the continuous symmetry on the fluid flow from a state space perspective.   
Subfigure \ref{fig:trajectories} (a) shows the trajectory of the DNS projected onto a 3-dimensional state-space for about 4160 advective time units matching the orbit's period. The projection $a_i=\left\langle\bm{u},\bm{e}_i\right\rangle$ was performed onto orthogonal unit vectors $\bm{e}_i$ created from linearly independent flow fields of the DNS. The two black dots indicate start and end point of the trajectory, which clearly lie separated from each other. This characterizes the stable period orbit as being relative. Figure \ref{fig:trajectories} (b) shows the trajectory of the symmetry-reduced dataset in the same state-space projection for the same interval in time. Here, start and end point fall together and the trajectory forms a closed loop. The low-momentum region crosses the simulation domain periodically, taking around 4000 advective time units for one cycle. During that cycle it changes its drift direction several times. 

\begin{figure}[t]
    \centering
    \includegraphics[width=\textwidth]{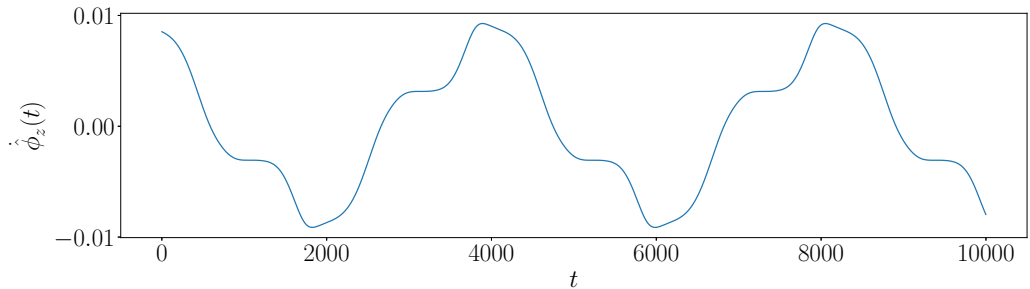}
    \caption{Slice phase velocity for the spanwise rotating ASBL sampled with time step $\delta=1$ obtained using the Equation (\ref{equ:reconstructionequz}) for a time window of 10,000 advective time units.}
    \label{fig:slice}
\end{figure}

To reduce the system by its continuous symmetry, we applied the first Fourier mode slicing technique to the DNS dataset using the template profile $f(y)=2.5T_2(y)$, where $T_2$ is the second Chebyshev polynomial of first kind. The template function was aligned with the spanwise direction. Figure \ref{fig:slice} illustrates the time dependence of the slice phase velocity $\dot{\hat{\phi}}_z$ calculated using the Equation (\ref{equ:reconstructionequz}). 
Subsequently, we applied the sparsity-promoting DMD to the original time series as well as the symmetry-reduced time series, in order to study the effect of the symmetry reduction. After some experimentation with the length and position of the sparsity-promoting DMD observation window, we set it to $T_w=4200$ advective time units starting from $t=0$. The snapshot separation in time was set to be $\delta t=95$ and the sampling interval to be $s=95$, leading to 43 snapshot pairs being processed by sparsity-promoting DMD. For the reconstruction, the sparsity-promoting DMD algorithm was set to select 9 dynamic modes. 
  
\begin{figure}[t]
\begin{minipage}[b]{0.5\textwidth}
\begin{subfigure}{\textwidth}
\subcaption{}
\includegraphics[width=\linewidth]{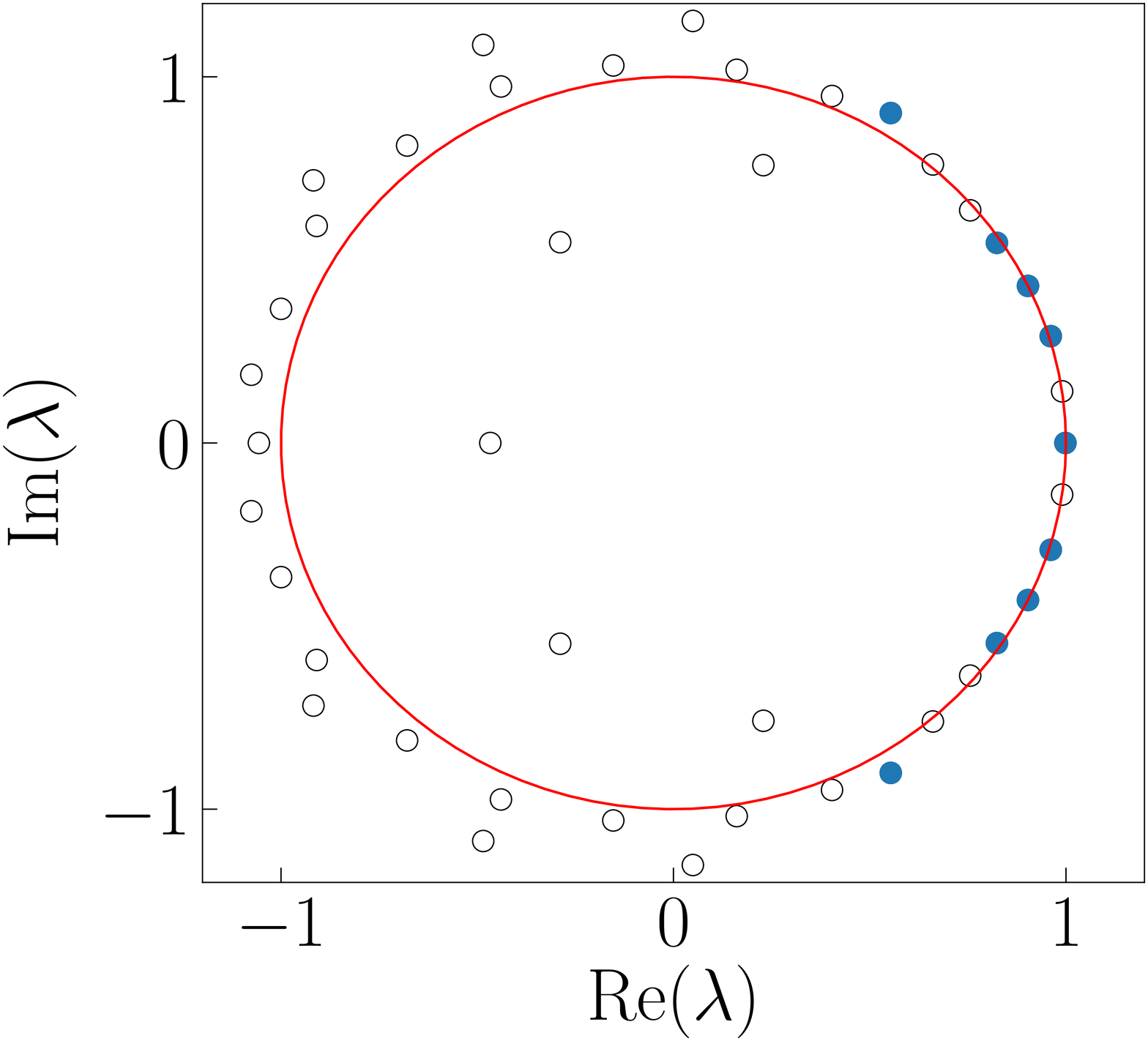}
\end{subfigure}\par
\end{minipage}
\begin{minipage}[b]{0.5\textwidth}
\begin{subfigure}{\textwidth}
\subcaption{}
\includegraphics[width=\linewidth]{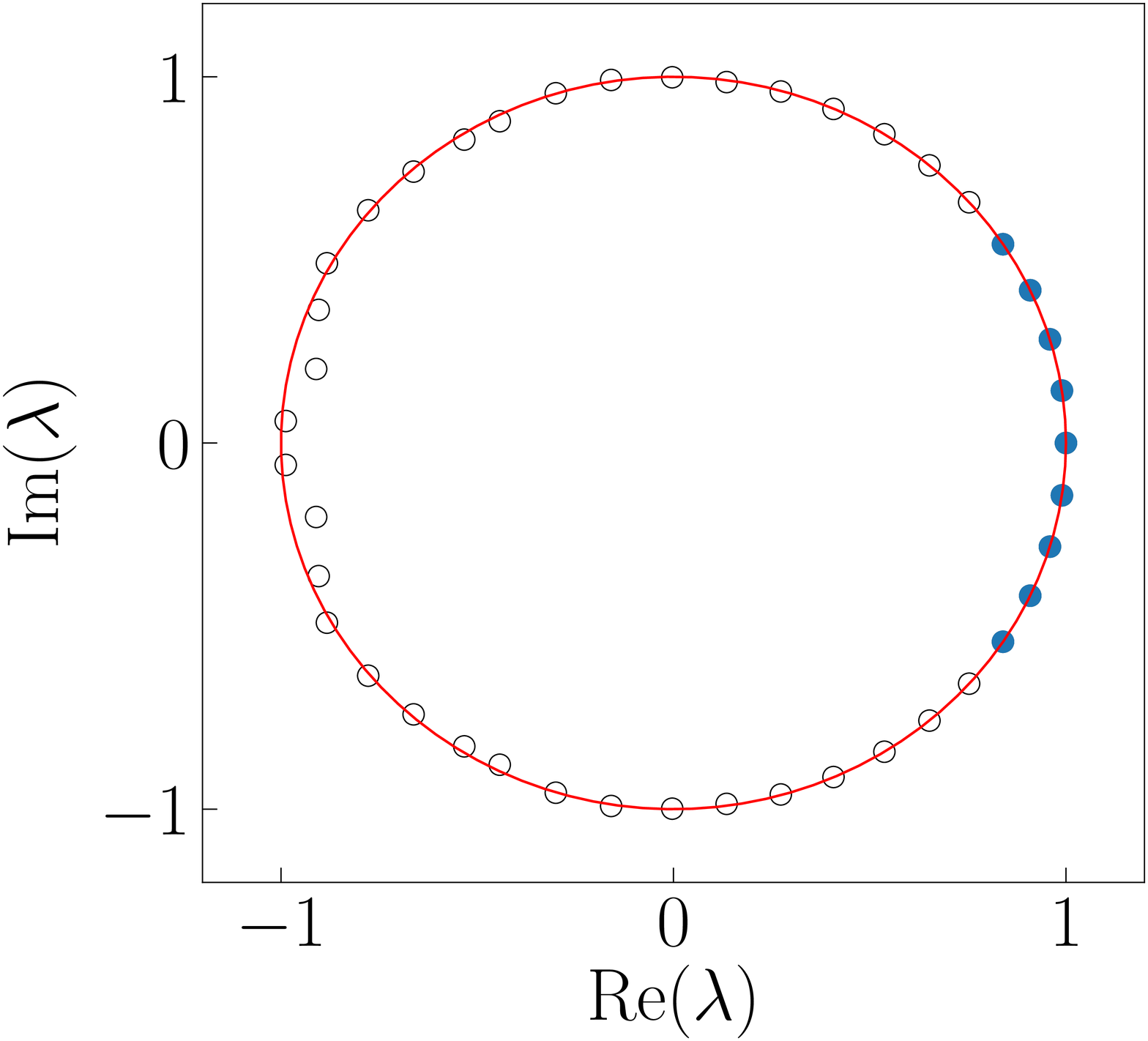}
\end{subfigure}
\end{minipage}
\quad
\begin{minipage}[b]{\textwidth}
\caption{Spectra obtained by using sparsity-promoting DMD for an SRPO in rotating ASBL. 
The eigenvalues in (a) are the result of applying sparsity-promoting DMD to the original time series and (b) to the symmetry-reduced time series. The filled circles indicate sparsity-promoting DMD eigenvalues used for the reconstruction.}\label{fig:spectra}
\end{minipage}
\end{figure}

Figure \ref{fig:spectra}(a,b) present the resulting DMD spectra for the original data in subfigure (a) and the symmetry-reduced data in subfigure (b). The dynamic modes selected by sparsity promotion are indicated with filled circles. Several observations can be made from the data. First, we observe that SRDMD is much better converged compared to DMD, as the eigenvalues in Figure~\ref{fig:spectra} (b) lie much almost all on the unit circle and are equidistantly distributed as they must be for a periodic orbit. In contrast, the eigenvalues shown in Figure~\ref{fig:spectra} (a) often lie outside the unit circle indicating exponentially growing dynamics or inside, indicating exponentially decaying dynamics. Neither is representative of the actual dynamics. Second, the sparsity-promoting algorithm selects the fundamental frequency and its first few harmonics out of the SRDMD eigenvalues, as expected.
For DMD however, the selected frequencies have gaps and fundamental frequency was not chosen for the reconstruction. Third, the period of the stable relative periodic orbit was estimated using the approximation for the fundamental frequency in (\ref{equ:Tg}) and its residual in (\ref{equ:residualTg}) giving $T_{\text{SRPO}}=4075.52\pm 0.0001$ for the non-symmetry reduced trajectory and $T_{\text{SRPO}}=4160.8\pm 2.4304\cdot 10^{-7}$ for the symmetry-reduced trajectory. These observations clearly indicate the effect of the time-dependent drift in the data and how the first Fourier mode slicing technique successfully removes the continuous symmetry from the system leading to much fit for $\bm{A}$ and therefore, to a much better decomposition and low-dimensional approximation of the data for SRDMD compared with DMD. 

\begin{figure}[t]
\begin{minipage}[b]{\textwidth}
\begin{subfigure}{\textwidth}
\subcaption{}
\includegraphics[width=\linewidth]{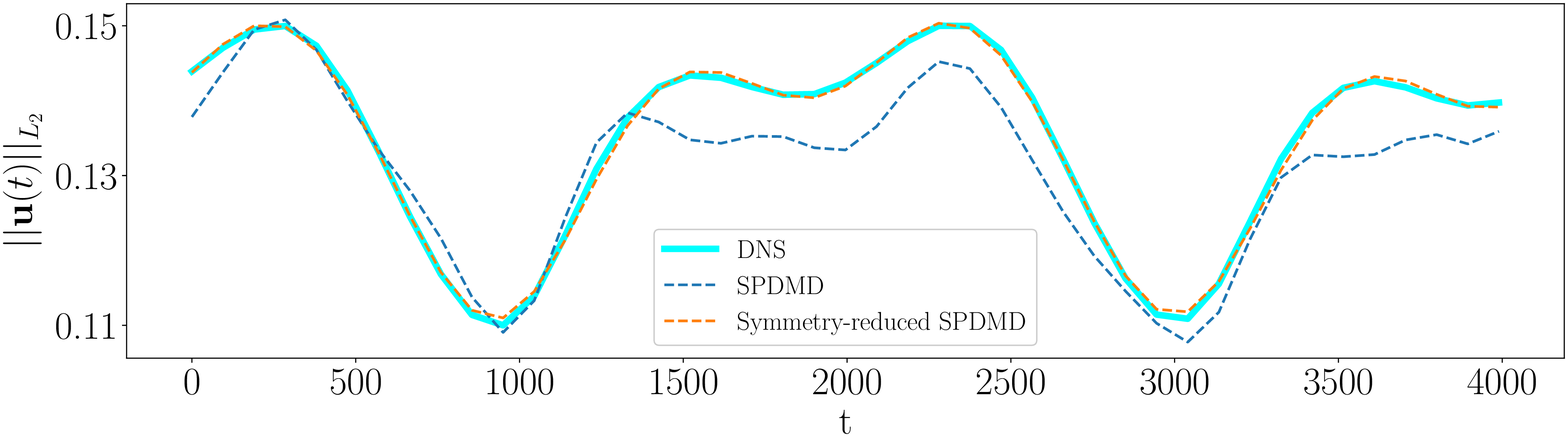}
\end{subfigure}\par
\end{minipage}
\begin{minipage}[b]{\textwidth}
\begin{subfigure}{\textwidth}
\subcaption{}
\includegraphics[width=\linewidth]{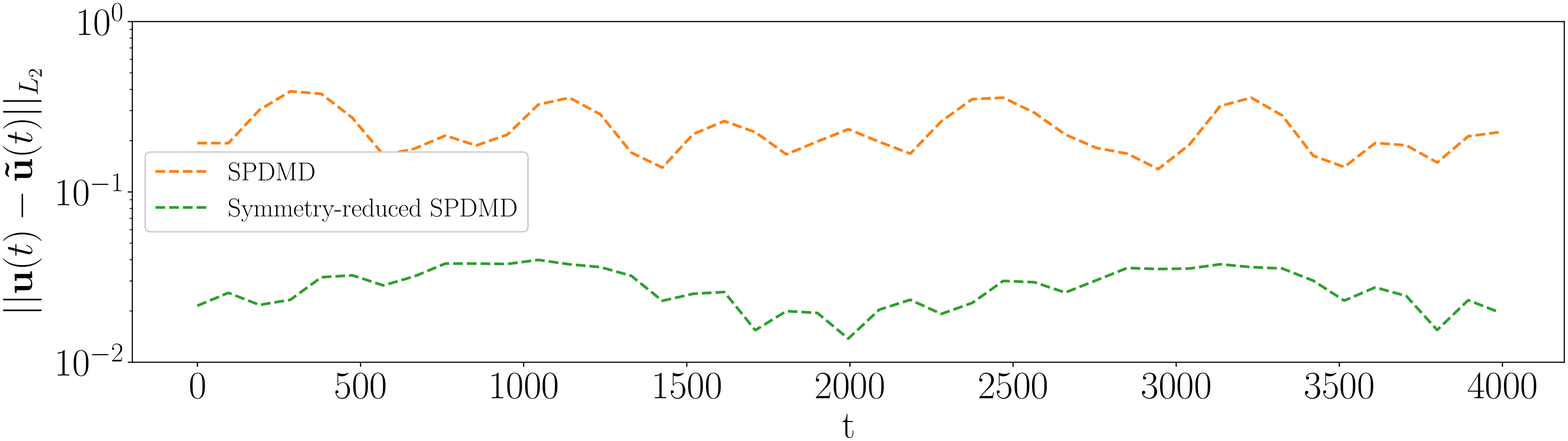}
\end{subfigure}
\end{minipage}
\quad
\begin{minipage}[b]{\textwidth}
\caption{Comparison of the DNS times series and the approximation with sparsity-promoting DMD using the $L_2$-distance of the flow fields as a function of time for the first 4000 advective time units. (a) Sparsity-promoting DMD applied to the non-symmetry reduced data set and (b) applied to the symmetry-reduced data set.}\label{fig:reconstruction}
\end{minipage}
\end{figure}

In summary, as expected \cite{characteritic_dmd}
the underlying continuous symmetry reduces the approximation quality of the low-dimensional representation. The loss of approximation quality is demonstrated qualitatively in Figure~\ref{fig:reconstruction}, where a comparison between the time series of 
$ \|\bm{u}\|_2$ for DNS data and the low-dimensional reconstructions obtained with sparsity-promoting DMD applied to the original dataset and the symmetry-reduced dataset are presented. Figure \ref{fig:reconstruction}(b) shows the $L_2$-distance between the snapshots of the original data and the respective reconstructions obtained with sparsity-promoting DMD before and after symmetry reduction as function of time. Both reconstructions capture the main features of the coherent structure, although the sparsity-promoting SRDMD clearly improves the approximation quality. With this result in mind, we now focus on turbulent ASBL. 

\section{Turbulent ASBL}\label{sec:2dasbl}

\begin{figure}[t]
\begin{minipage}[b]{0.45\textwidth}
\begin{subfigure}{\textwidth}
\subcaption{}
\includegraphics[width=\linewidth]{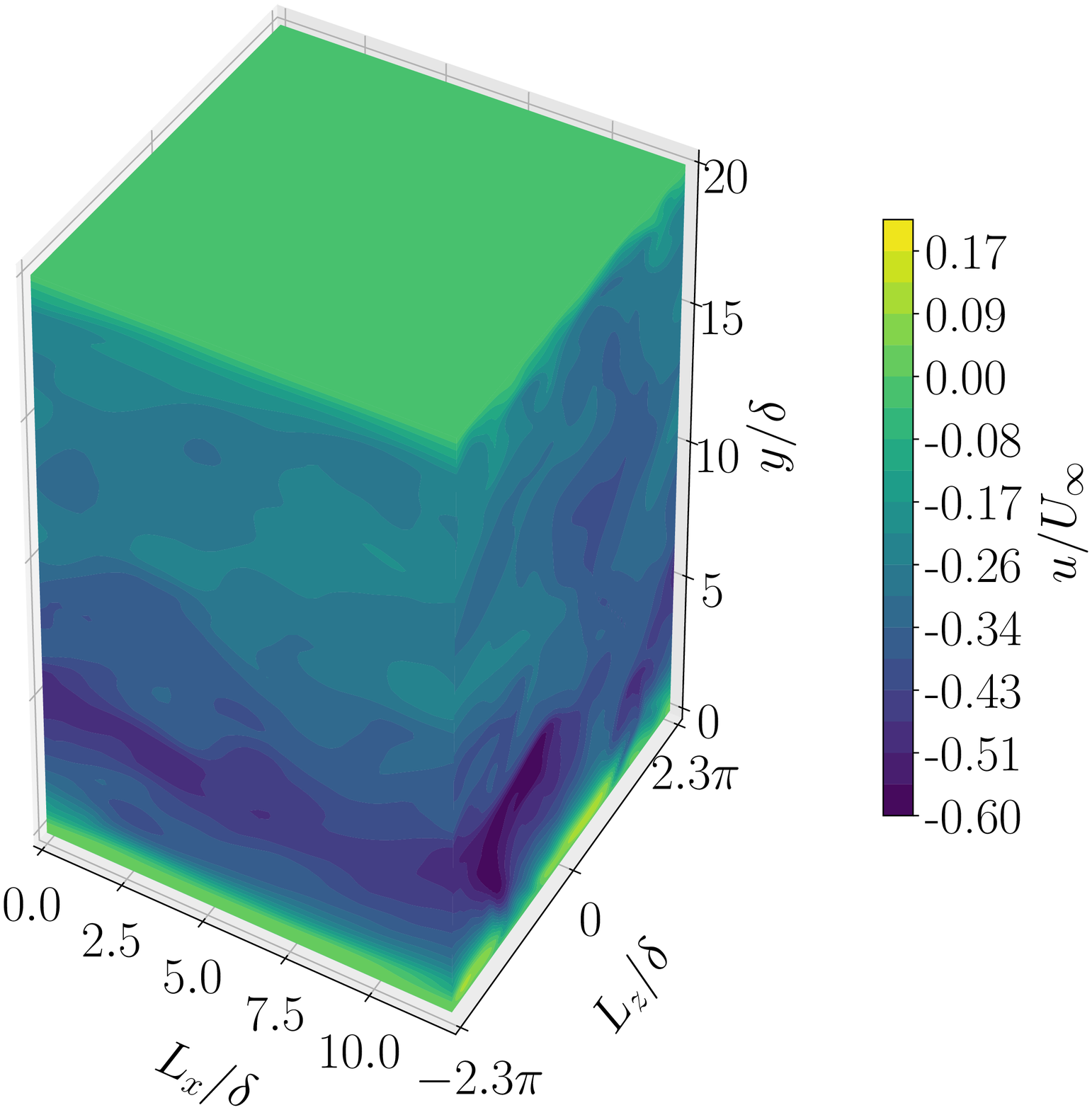}
\end{subfigure}\par
\end{minipage}
\begin{minipage}[b]{0.45\textwidth}
\begin{subfigure}{\textwidth}
\subcaption{}
\includegraphics[width=\linewidth]{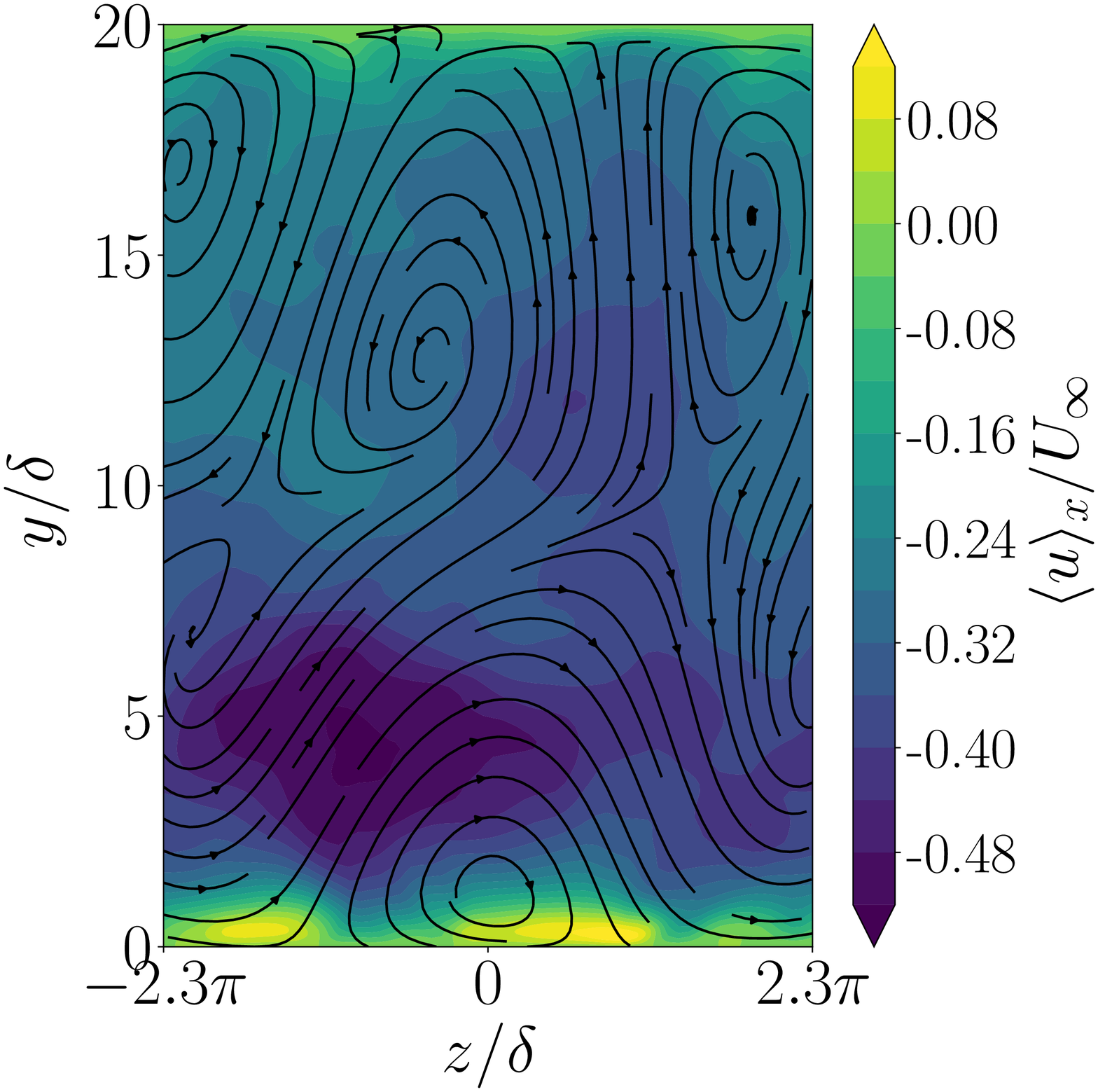}
\end{subfigure}
\end{minipage}
\quad
\begin{minipage}[b]{\textwidth}
\caption{(a) 3d streamwise velocity contour plot of the snapshot at $t=0\delta/U_{\infty}$. The colour code indicates the deviation of the streamwise velocity from the laminar profile. (b) shows (a) averaged in streamwise direction. The lines and arrows indicate the $y$-$z$ streamfunction.}\label{fig:3dasbl}
\end{minipage}
\end{figure}

\begin{table}[h]
    \begin{center}
        \begin{tabular}{||c c c c c c c c c c||}
            \hline
            Re & Re$_\tau$ & $h/\delta$ & $L_z/\delta$ & $L_x/\delta$ & $N_x$ & $N_y$ & $N_z$ & $\Delta x^+$ & $\Delta z^+$ \\
            \hline
            1000 & 327 & 20 & $4.6\pi$ & $4\pi$ & 64 & 161 & 96 & 5.1 & 3.9 \\
            \hline
        \end{tabular}
    \end{center}
\caption{Numerical details. Shown are the Reynolds number $Re$, the friction Reynolds number $\text{Re}_{\tau}=u_{\tau}\delta_{0.99}/\nu$ based on the friction velocity $u_{\tau}$, the kinematic viscosity $\nu$ and the boundary layer thickness $\delta_{0.99}\approx 19.98$. Given are the length $L_x$, width $L_z$ and height $h$ of the computational domain with grid points $N_x$, $N_z$ and $N_y$. The grid resolution $\Delta x^+$ and $\Delta z^+$ in wall units considers $2/3^{\text{rds}}$ dealiasing in streamwise and spanwise direction.}
\label{tab:tasbl}
\end{table}

\begin{figure}[t]
\begin{minipage}[b]{0.292\textwidth}
\begin{subfigure}{\textwidth}
\subcaption{}
\includegraphics[width=\linewidth]{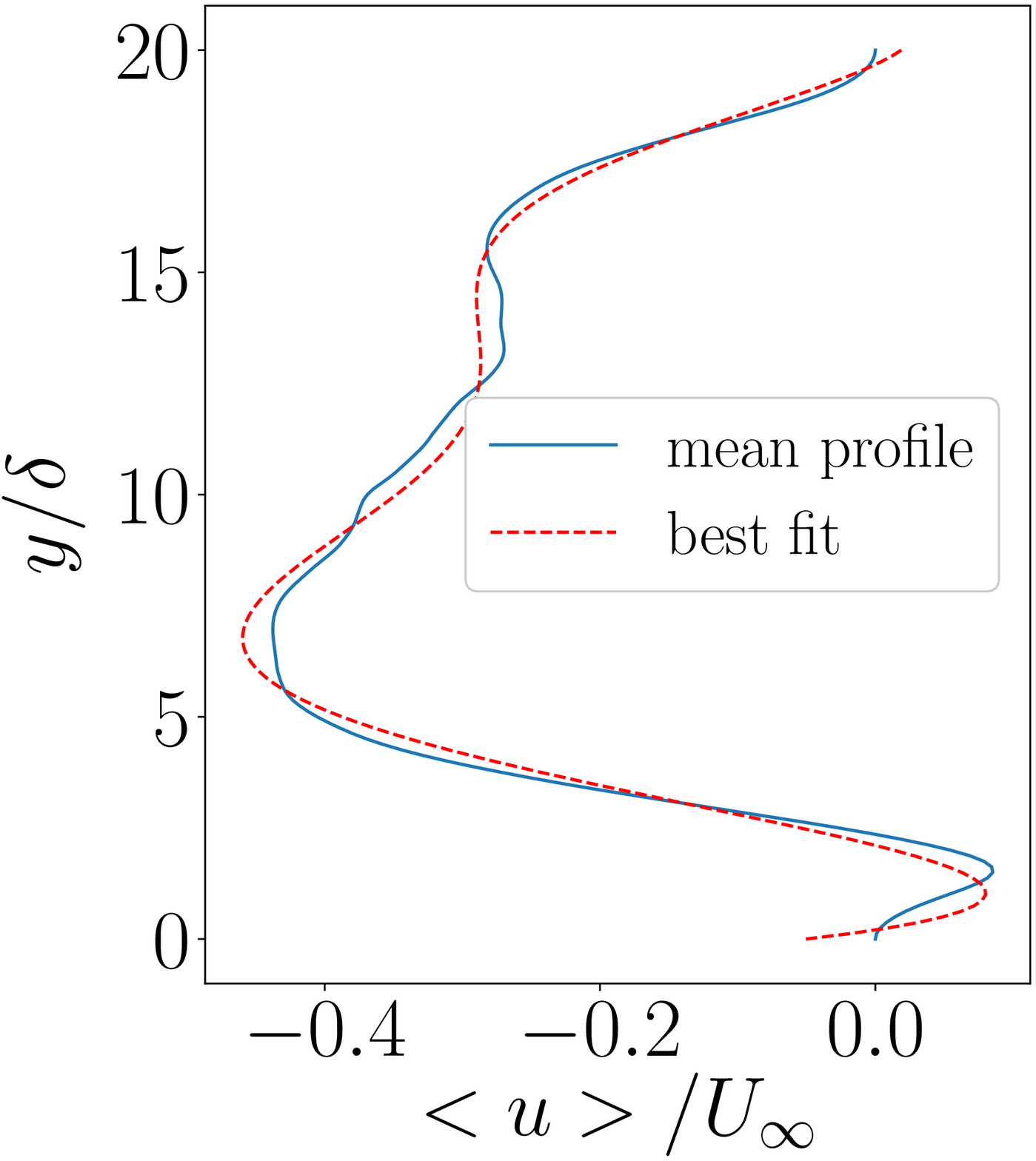}
\end{subfigure}\par
\end{minipage}
\begin{minipage}[b]{0.335\textwidth}
\begin{subfigure}{\textwidth}
\subcaption{}
\includegraphics[width=\linewidth]{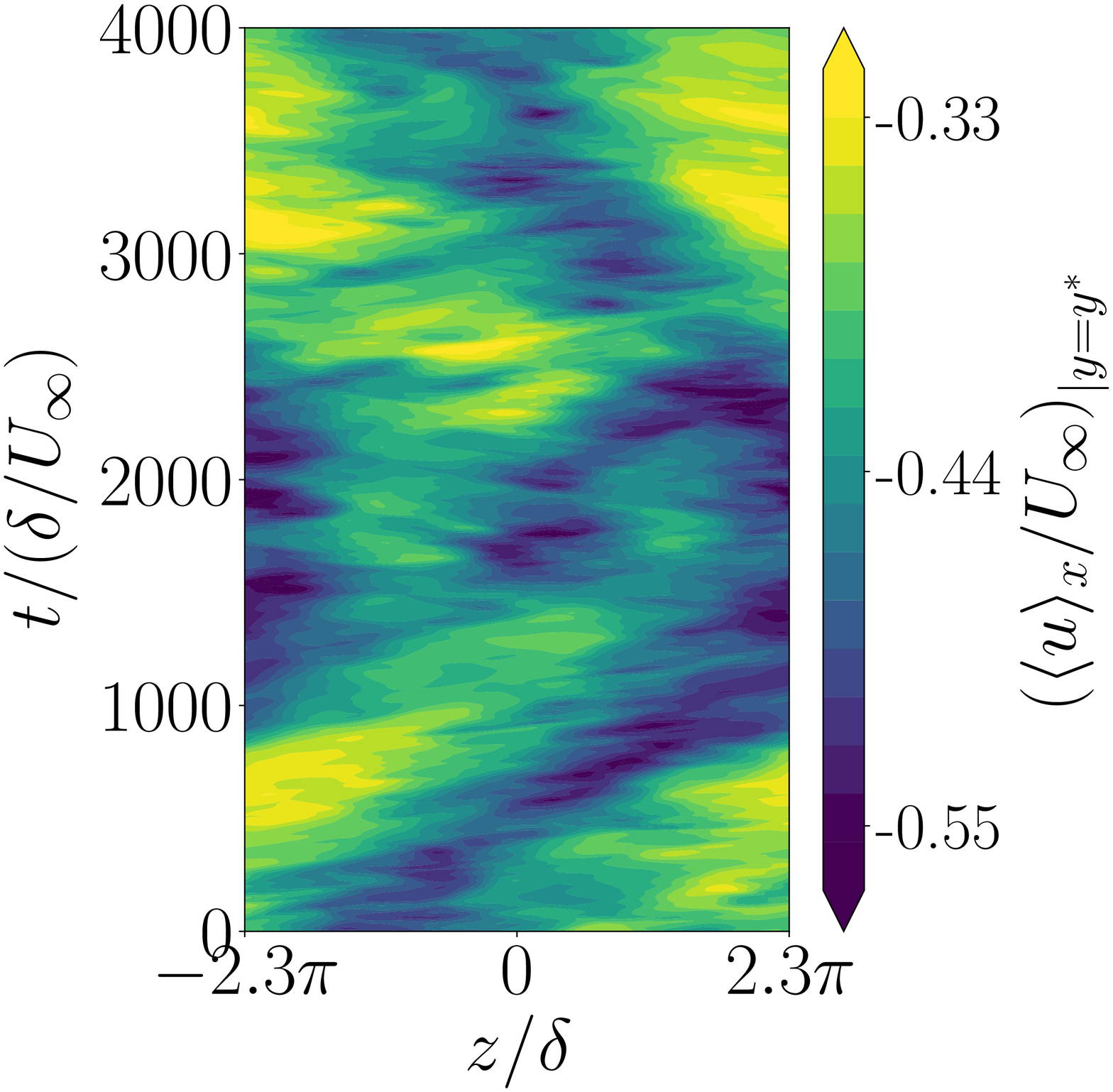}
\end{subfigure}
\end{minipage}
\begin{minipage}[b]{0.335\textwidth}
\begin{subfigure}{\textwidth}
\subcaption{}
\includegraphics[width=\linewidth]{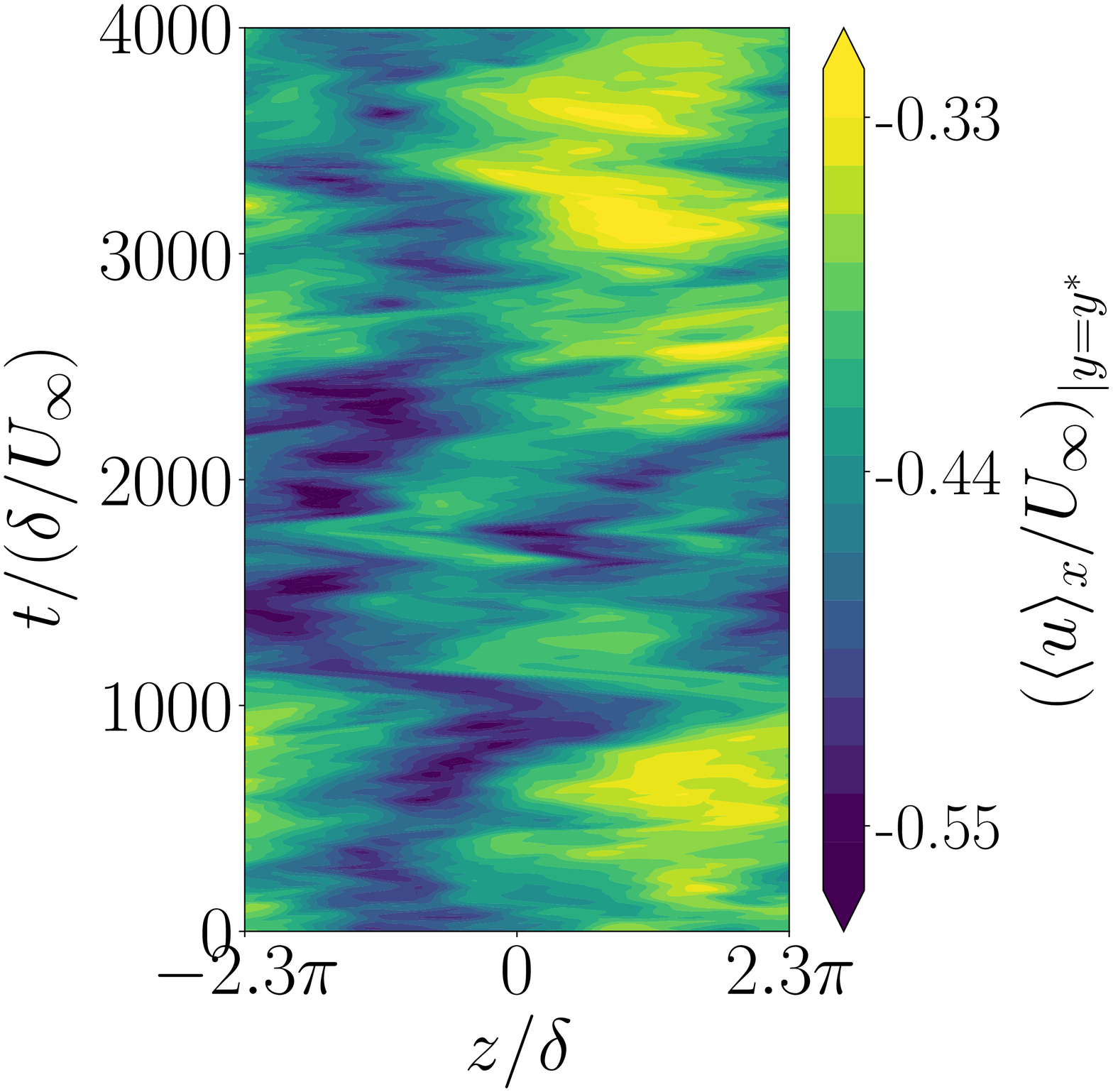}
\end{subfigure}
\end{minipage}
\quad
\begin{minipage}[b]{\textwidth}
\caption{(a) Mean velocity profile in streamwise direction averaged over the first approx. 300 snapshots indicated by the solid blue line. The dashed red line shows the best fit for a Chebyshev polynomial-based function. (b) Spatio-temporal structure of the trajectory for 4000 advective time units as a function of spanwise direction and time. The colour code indicates the streamwise averaged deviation of the streamwise velocity from the laminar profile along the horizontal line in spanwise direction at wall-normal height $y/\delta\approx 3$ for the original data set and (c) for the symmetry-reduced data set.}\label{fig:ucontour-nonsymred}
\end{minipage}
\end{figure}

\begin{figure}[t]
\begin{minipage}[b]{0.24\textwidth}
\begin{subfigure}{\textwidth}
\subcaption{}
\includegraphics[width=\linewidth]{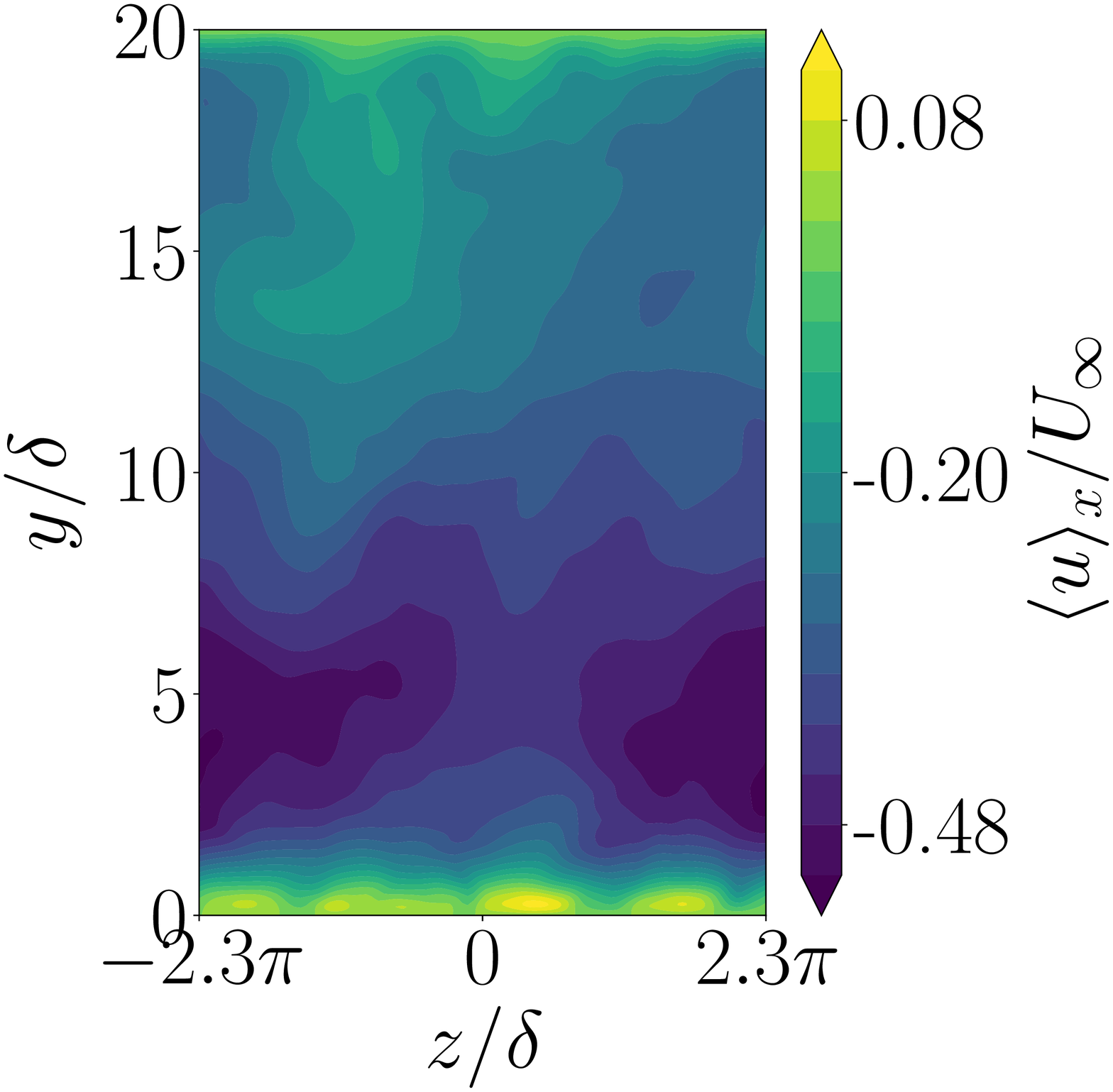}
\end{subfigure}\par
\end{minipage}
\begin{minipage}[b]{0.24\textwidth}
\begin{subfigure}{\textwidth}
\subcaption{}
\includegraphics[width=\linewidth]{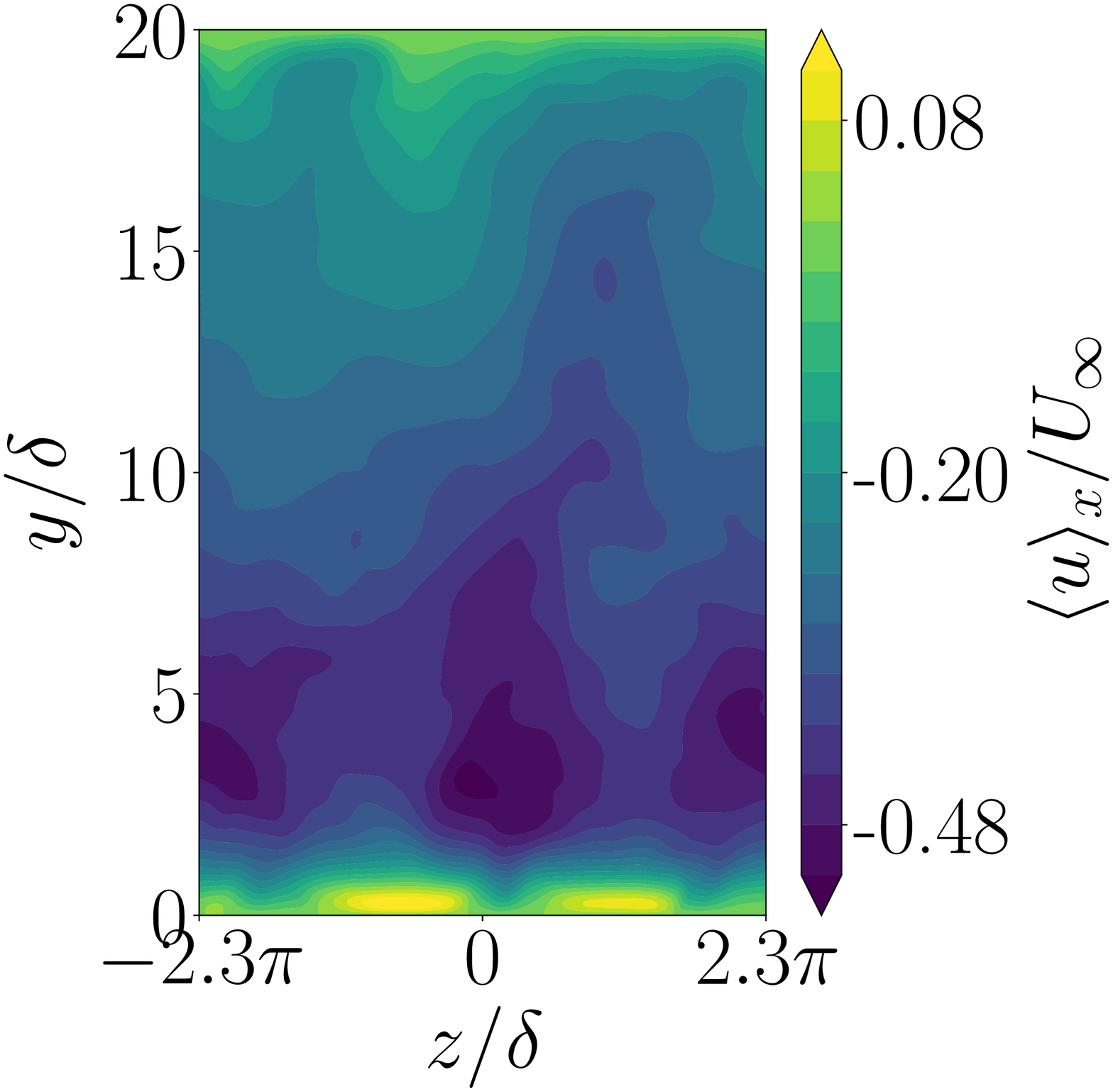}
\end{subfigure}
\end{minipage}
\begin{minipage}[b]{0.24\textwidth}
\begin{subfigure}{\textwidth}
\subcaption{}
\includegraphics[width=\linewidth]{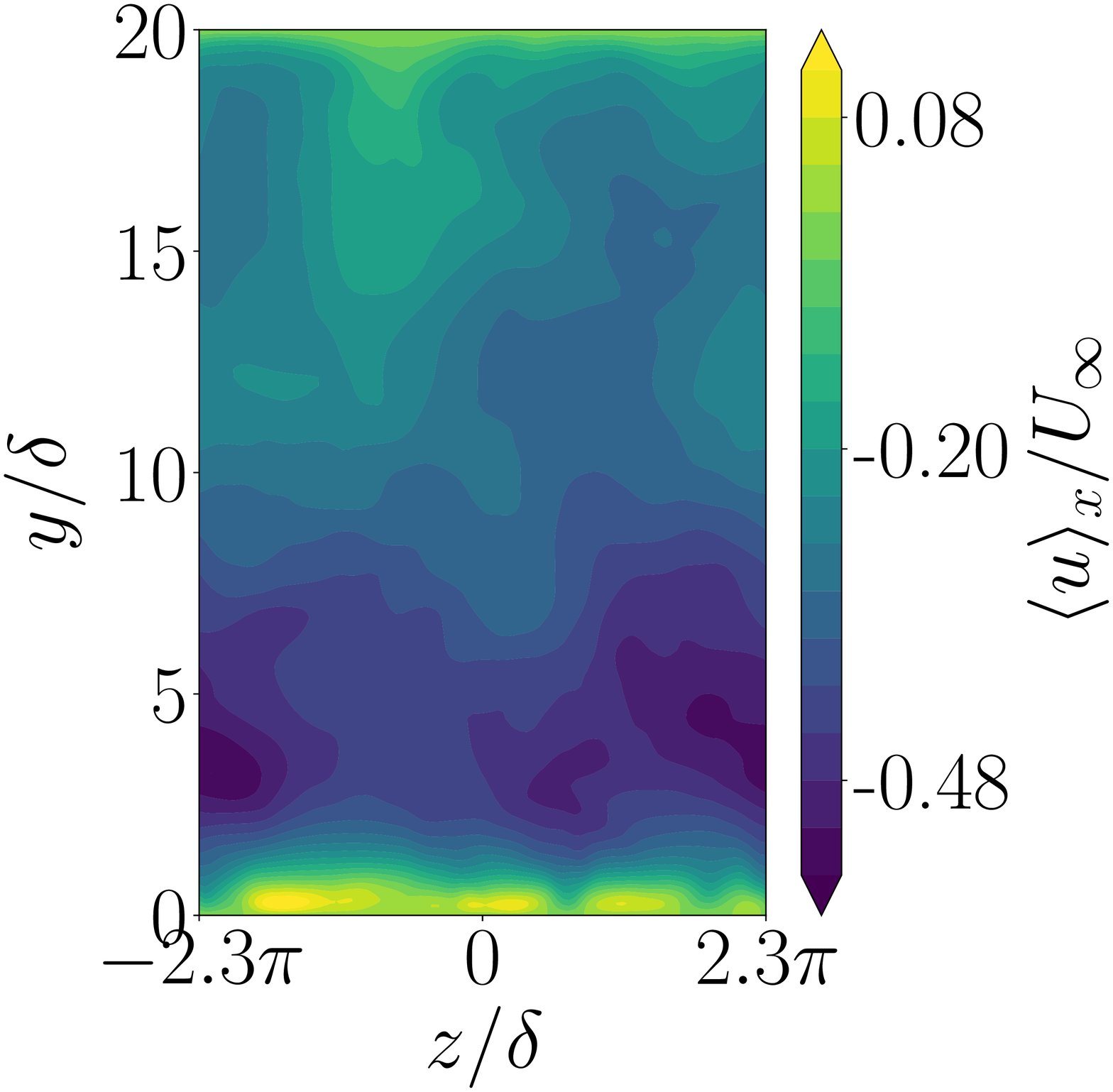}
\end{subfigure}
\end{minipage}
\begin{minipage}[b]{0.24\textwidth}
\begin{subfigure}{\textwidth}
\subcaption{}
\includegraphics[width=\linewidth]{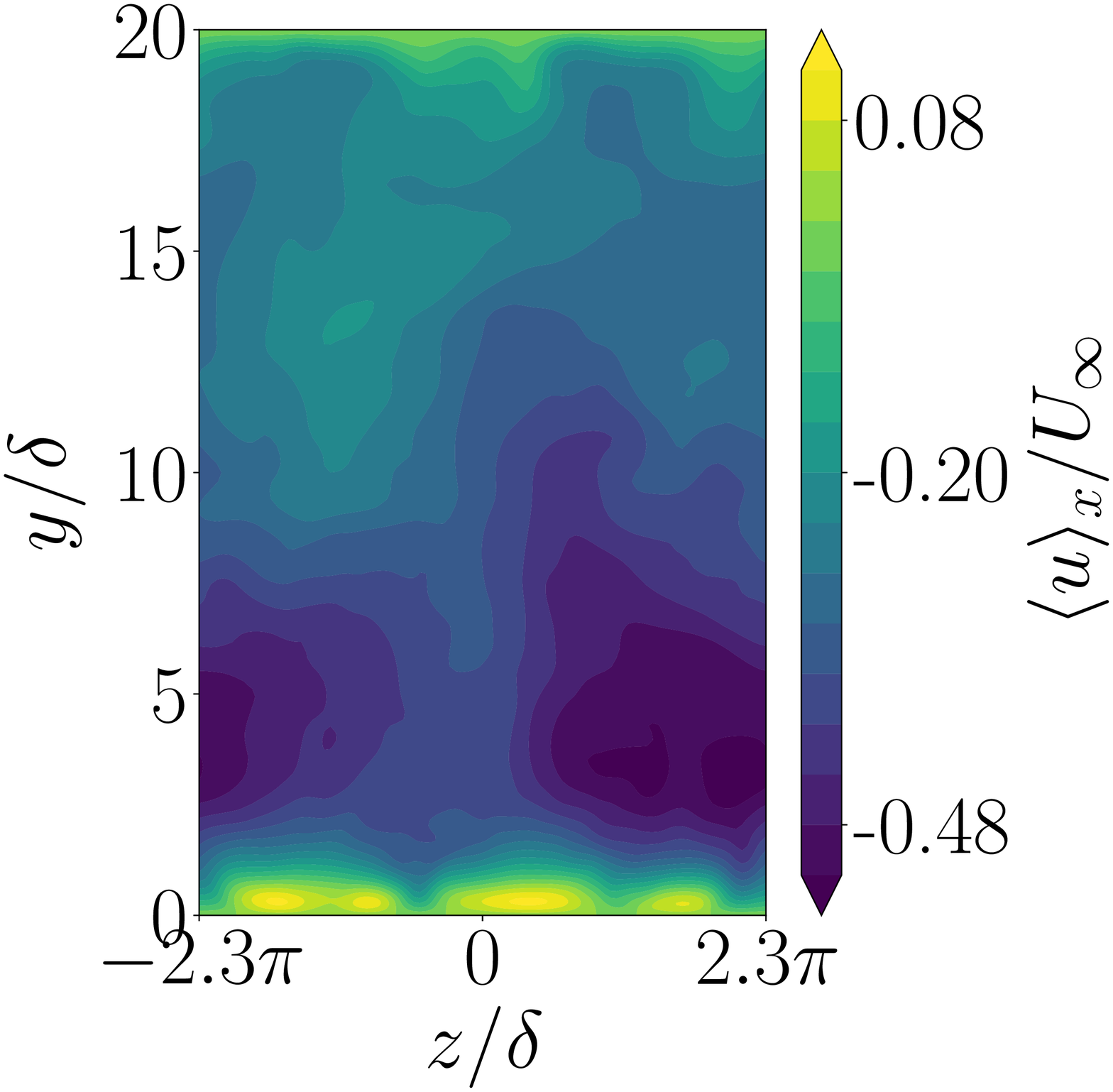}
\end{subfigure}
\end{minipage}
\quad
\begin{minipage}[b]{\textwidth}
\caption{The occurrence, drift and merging process of a low-momentum region. Shown are the velocity contour plots of DNS snapshots at time (a) $t=1300\delta/U_{\infty}$, (b) $t=1680\delta/U_{\infty}$, (c) $t=2100\delta/U_{\infty}$ and (d) $t=2250\delta/U_{\infty}$. The colour code indicates the streamwise-averaged deviation of the streamwise velocity from the laminar profile.}\label{fig:ucontour-split}
\end{minipage}
\end{figure}

In this section, we discuss the application of the first Fourier mode slicing technique for the reduction of continuous symmetries and the subsequent sparsity-promoting SRDMD for obtaining a low-dimensional representation of turbulent ASBL. The Reynolds number was set to be $\text{Re}=1000$ with a time averaged friction Reynolds number of $\text{Re}_{\tau}=327$. The computational domain is $L_x/\delta\times h/\delta\times L_z/\delta=4\pi\times 20\times 4.6\pi$ with a resolution of $64\times 161\times 96$ modes. Further details such as resolution in wall units are given in table \ref{tab:tasbl}.

We extend on preliminary work on large-scale coherence in the ASBL \cite{streamingdmd}, carried out on two-dimensional data obtained by streamwise averaging.  
Firstly, the analysis is now carried out on the full three-dimensional flow fields, hence an additional shift symmetry in streamwise direction must be accounted for. Second, we remove dynamically irrelevant drifts using SRDMD~\cite{marensi_yalniz_hof_budanur_2023} instead of working in a co-moving frame. Most importantly, the spanwise drift in the present dataset is time-dependent, whereas the previous analysis was carried out on a subsection of the trajectory, in which the low-momentum region was drifting in spanwise direction at an approximately constant velocity. 

Figure \ref{fig:3dasbl} shows a three-dimensional streamwise-velocity contour plot of a snapshot at time $t=0\delta/U_{\infty}$. The colour code indicates the deviation of the streamwise velocity from the laminar profile, $u$. We followed the turbulent trajectory for about 4000 advective time units and observed a large-scale low-momentum region localised in spanwise direction with its center at the wall-normal height $y/\delta\approx 3$. This coherent structure is accompanied by high-speed streaks near the bottom wall, shown in Figure~\ref{fig:3dasbl} (b), which presents the streamwise-averaged flow field the same instance in time. The colour coding corresponds to the streamwise averaged deviation of the streamwise velocity from the laminar profile, $\langle u \rangle_x$ and the streamlines to the streamwise-averaged cross-flow, $(\langle v\rangle_x, \langle w \rangle_x)$.  The flow has a fast drift in streamwise and a slow drift in spanwise direction, leading the large-scale low-momentum zone to move transversely through the computational domain. 

Figures \ref{fig:ucontour-nonsymred}(b) presents the spatio-temporal structure of $\langle u \rangle_x$ as a function of the spanwise direction and time, measured at wall-normal height $y/\delta\approx 3$ for the original data. The clearly discernible oblique stripe pattern indicates the drift of the large-scale low-momentum region through the entire computational domain in spanwise direction within about $t\approx 2000\delta/U_{\infty}$ advective time units, after which it turns its drift direction to return back to its starting position after additional $t\approx 2000\delta/U_{\infty}$ advective time units. As can be seen, the large-scale low-momentum zone persists over the full trajectory despite its variation in intensity. Furthermore, we observe the appearance of a second low-momentum zone at about $t\approx 1600\delta/U_{\infty}$ and a spanwise coordinate $z/\delta\approx 0$, which drifts in positive spanwise direction taking around 1000 advective time units to finally merge with the dominant low-momentum region. This is supported by the DNS snapshots shown in Figure~\ref{fig:ucontour-split}, which illustrate this event. 

The drift of the primary low-momentum region is dynamically irrelevant. Therefore, we applied the first Fourier mode slicing technique to remove the continuous symmetries from the data. First, we reduced the continuous symmetry in streamwise direction with a template function aligned to the streamwise direction. Subsequenty, we reduced the continuous symmetry in spanwise direction with the template function aligned to the spanwise direction. The success of the slicing technique depends strongly on how the template profile fits to the velocity profiles of the snapshots over the whole trajectory. In both cases, we chose the template profile to be a fit to the mean velocity profile as shown in Figure~\ref{fig:ucontour-nonsymred}(a), which is justified by the low variance of the velocity from the mean profile over time. The snapshots were sampled with $\delta t =1\delta/U_{\infty}$ over the full trajectory of 4000 advective time units. 

\begin{figure}[t]
\hfill
\begin{minipage}[b]{\textwidth}
\begin{subfigure}{\textwidth}
\subcaption{}
\includegraphics[width=\linewidth]{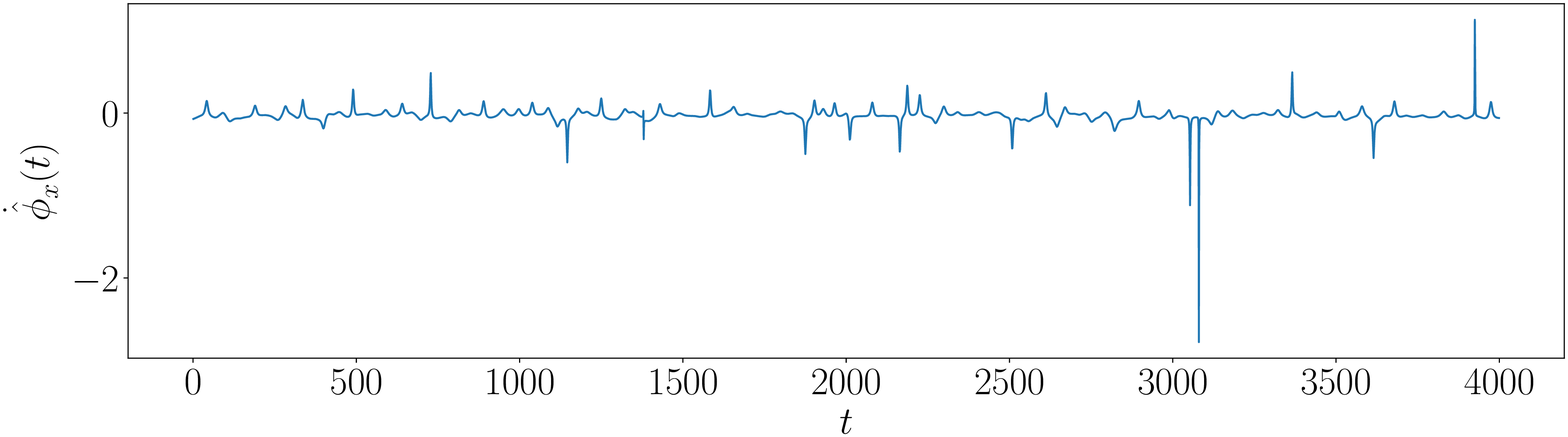}
\end{subfigure}\par
\begin{subfigure}{\textwidth}
\subcaption{}
\includegraphics[width=\linewidth]{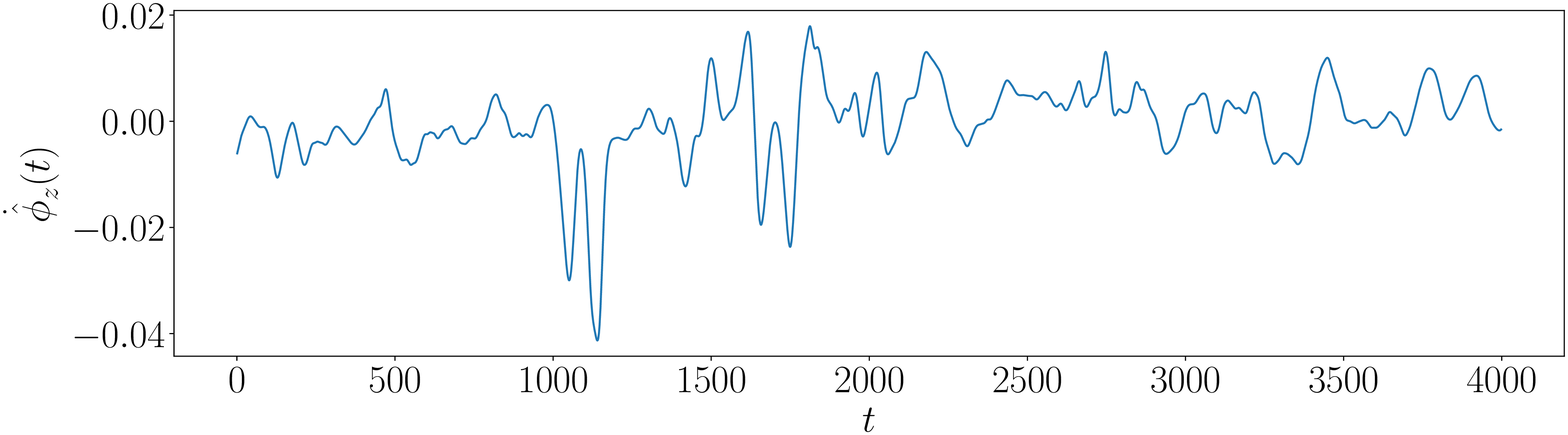}
\end{subfigure}
\end{minipage}
\quad
\begin{minipage}[b]{\textwidth}
\caption{Phase velocities for (a) the streamwise and (b) spanwise drift as a function of time sampled with time step $\delta t=1$, obtained by calculating the reconstruction equations (\ref{equ:reconstructionequx}-\ref{equ:reconstructionequz}).}\label{fig:tasbl_phasevel}
\end{minipage}
\end{figure}

Figure \ref{fig:tasbl_phasevel} shows the resulting phase velocities as functions of time for both directions obtained with the first Fourier mode slicing technique. 
Both phase velocities are well defined over the full observation window but show intervals of fast phase changes in time indicating a complicated transverse drift. The space-time diagram presented in Figure~\ref{fig:ucontour-nonsymred}(c) shows the result of the symmetry reduction with the spatio-temporal structure of the large-scale low-momentum region for the trajectory being restricted to the slice. Compared with the spatio-temporal diagram showing the original data, Figure~\ref{fig:ucontour-nonsymred}(b), it can be seen that the oblique pattern has been replaced by a vertical one indicating that the spanwise drift has been removed up to small fluctuations. We point out that the first Fourier mode slice method fixes only the largest scales in the system, while smaller scales are free to drift (N.B. Bundanur, private communication). 

\begin{figure}[t]
\begin{minipage}[b]{0.32\textwidth}
\begin{subfigure}{\textwidth}
\subcaption{}
\includegraphics[width=\linewidth]{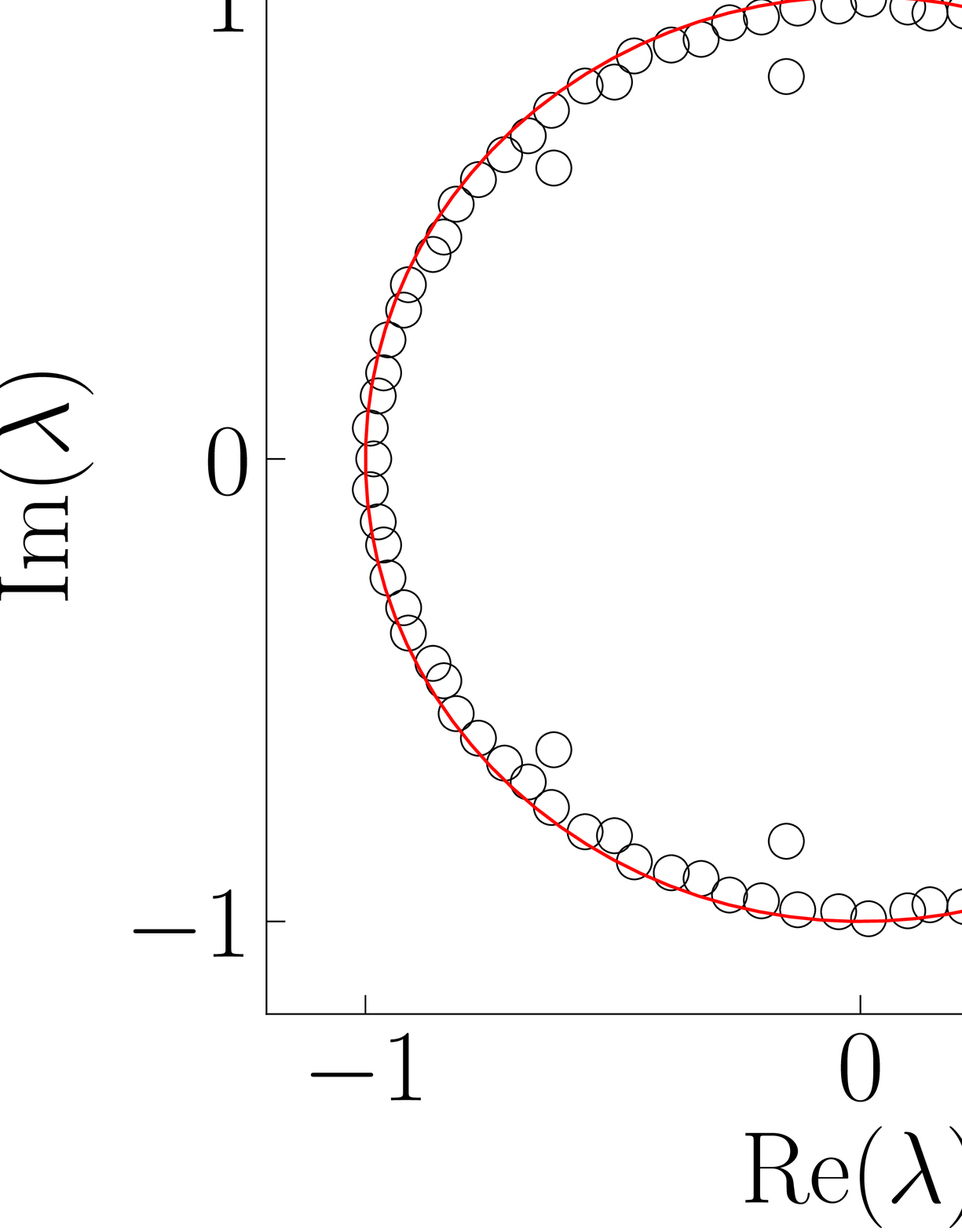}
\end{subfigure}\par
\end{minipage}
\begin{minipage}[b]{0.32\textwidth}
\begin{subfigure}{\textwidth}
\subcaption{}
\includegraphics[width=\linewidth]{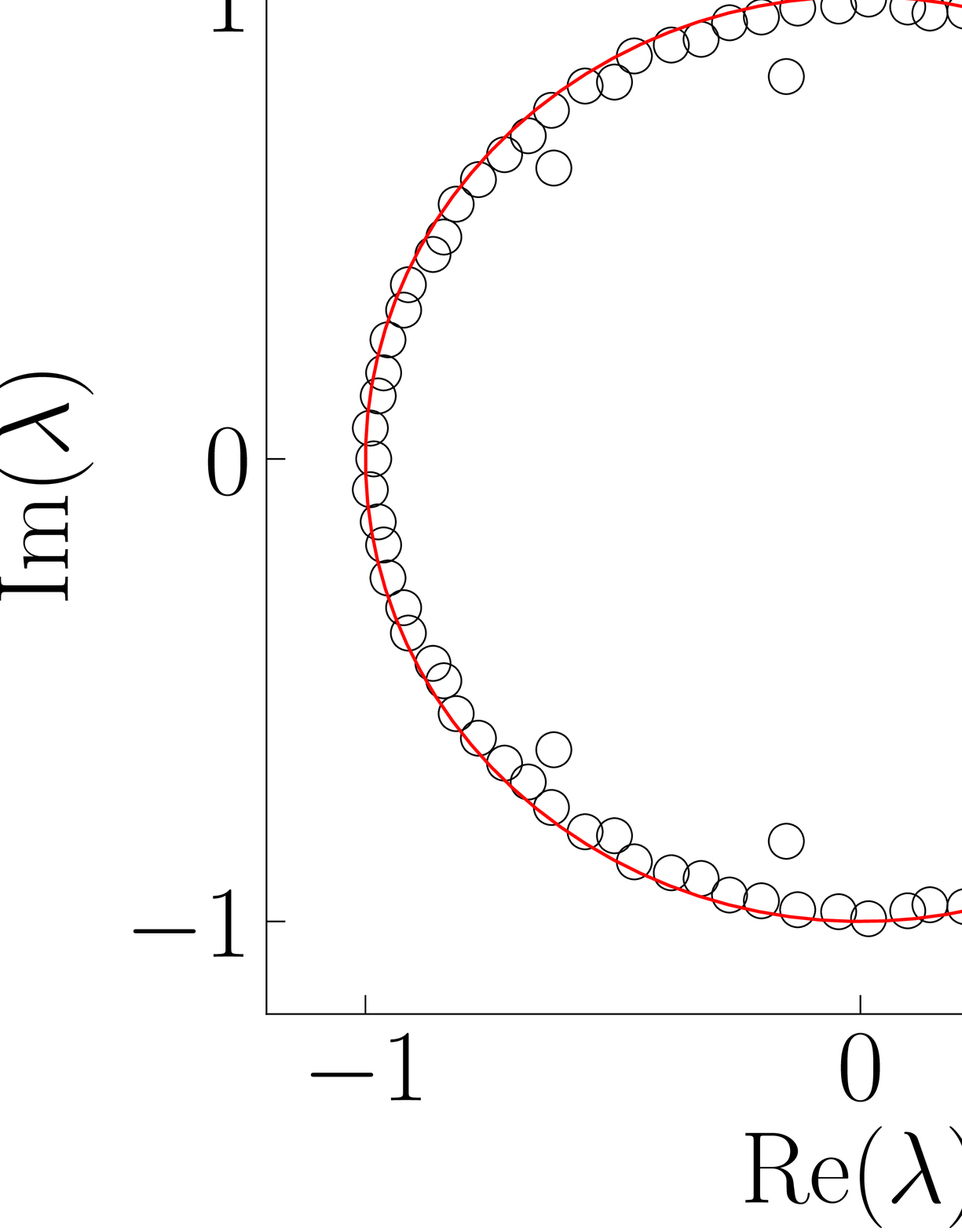}
\end{subfigure}
\end{minipage}
\begin{minipage}[b]{0.32\textwidth}
\begin{subfigure}{\textwidth}
\subcaption{}
\includegraphics[width=\linewidth]{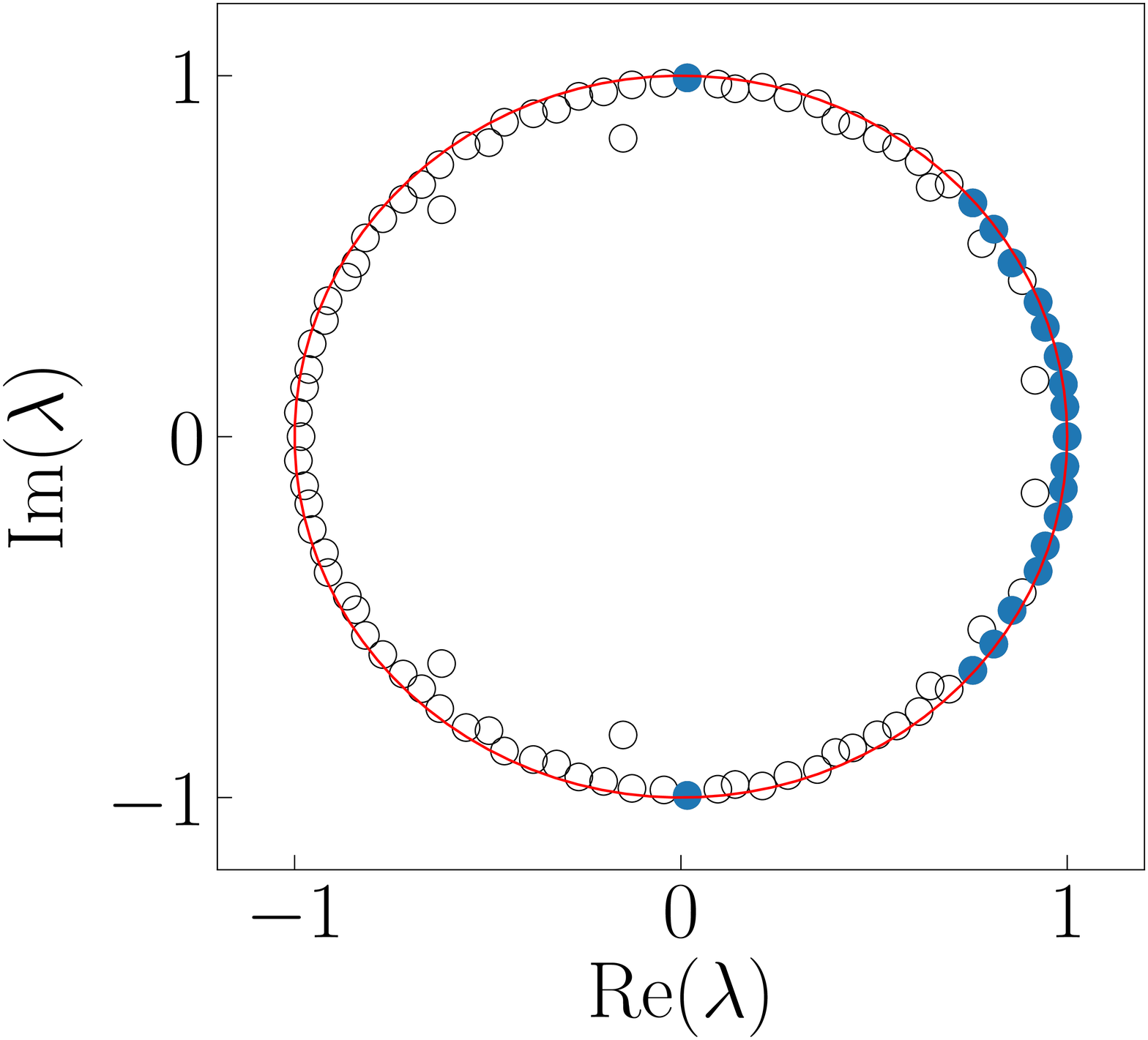}
\end{subfigure}
\end{minipage}
\quad
\begin{minipage}[b]{\textwidth}
\caption{Spectra obtained by applying sparsity-promoting DMD to the symmetry-reduced data set. The filled circles indicate eigenvalues corresponding to (a) seven, (b) thirteen and (c) nineteen dynamic modes selected for the low-dimensional reconstruction. The red line indicates the unit circle.}\label{fig:eigenvalues-spdmd-asbl}
\end{minipage}
\end{figure}

The objective of the subsequent sparsity-promoting DMD was to approximate the symmetry-reduced data set and to capture its essential features in a low-dimensional representation. Based on a convergence study considering the distance of the DMD frequencies from the unit circle, we chose a separation time of $\delta t=40\delta/U_{\infty}$ between the snapshot pairs $\bm{u}(t_j)$ and $\bm{u}(t_j+\delta t)$, as well as a sampling interval of $\Delta s=t_{j+1}-t_j=40\delta/U_{\infty}$. This led to 98 snapshot pairs processed by the sparsity-promoting DMD. Figure \ref{fig:eigenvalues-spdmd-asbl}(a-c) shows the spectra obtained by SRDMD and the eigenvalues selected by sparsity-promoting resulting in seven, 13 and 19 dynamic modes, respectively. The filled circles in subfigures (a), (b) and (c) indicate the eigenvalues corresponding to the dynamic modes used in the reconstruction.

\begin{figure}[t]
\begin{minipage}[b]{0.32\textwidth}
\begin{subfigure}{\textwidth}
\subcaption{}
\includegraphics[width=\linewidth]{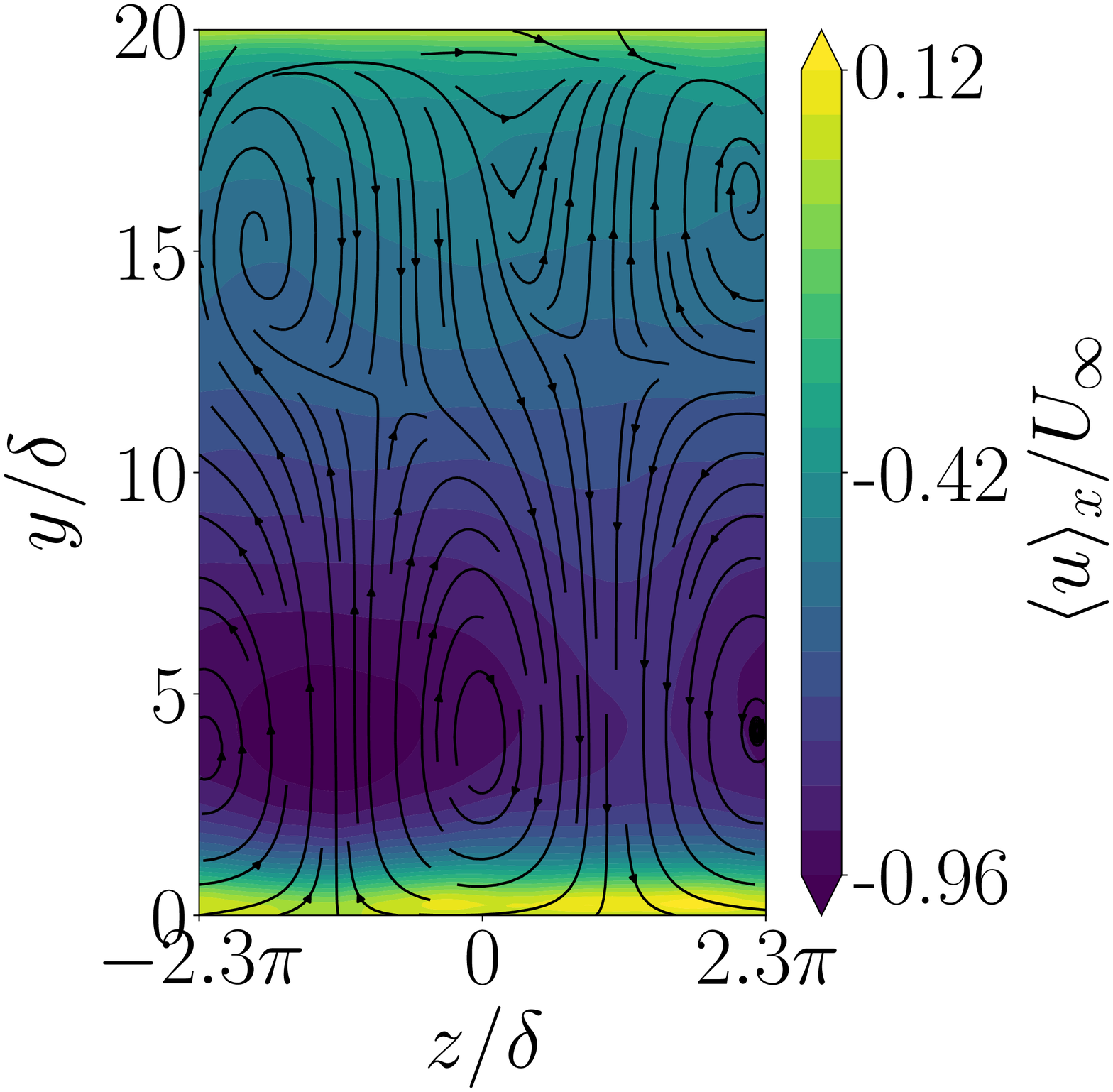}
\end{subfigure}\par
\end{minipage}
\begin{minipage}[b]{0.32\textwidth}
\begin{subfigure}{\textwidth}
\subcaption{}
\includegraphics[width=\linewidth]{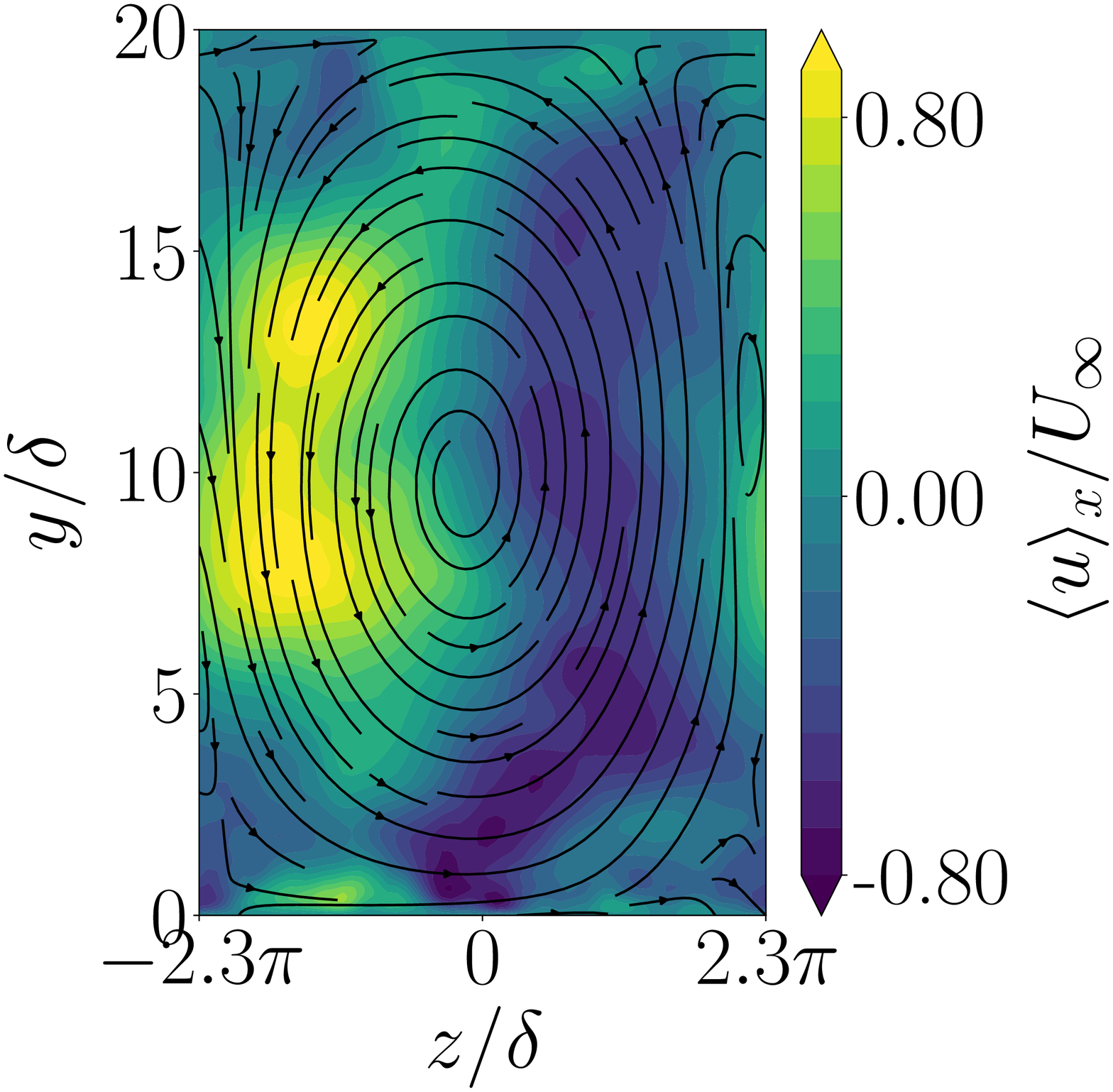}
\end{subfigure}
\end{minipage}
\begin{minipage}[b]{0.32\textwidth}
\begin{subfigure}{\textwidth}
\subcaption{}
\includegraphics[width=\linewidth]{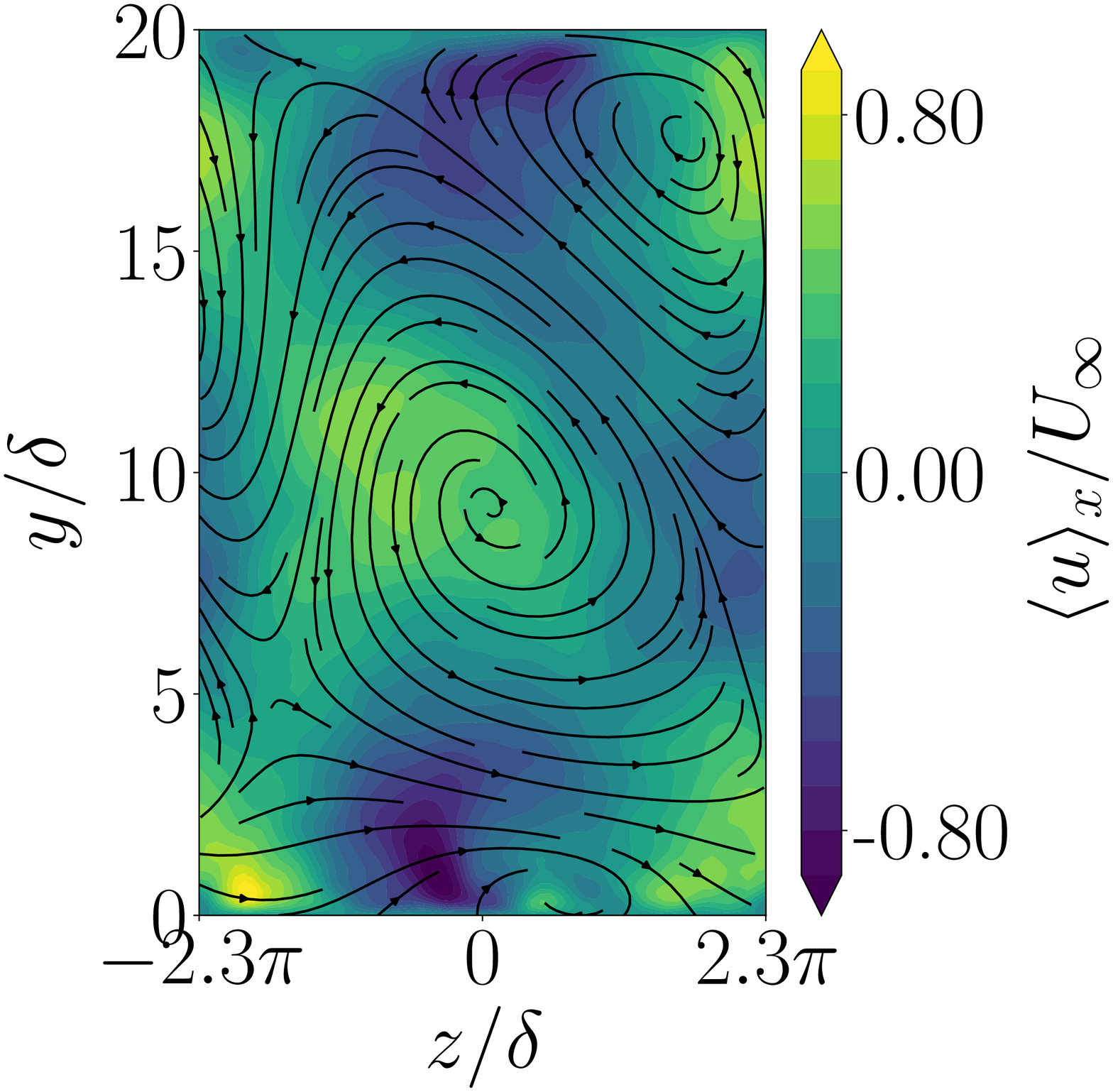}
\end{subfigure}
\end{minipage}
\begin{minipage}[b]{0.32\textwidth}
\begin{subfigure}{\textwidth}
\subcaption{}
\includegraphics[width=\linewidth]{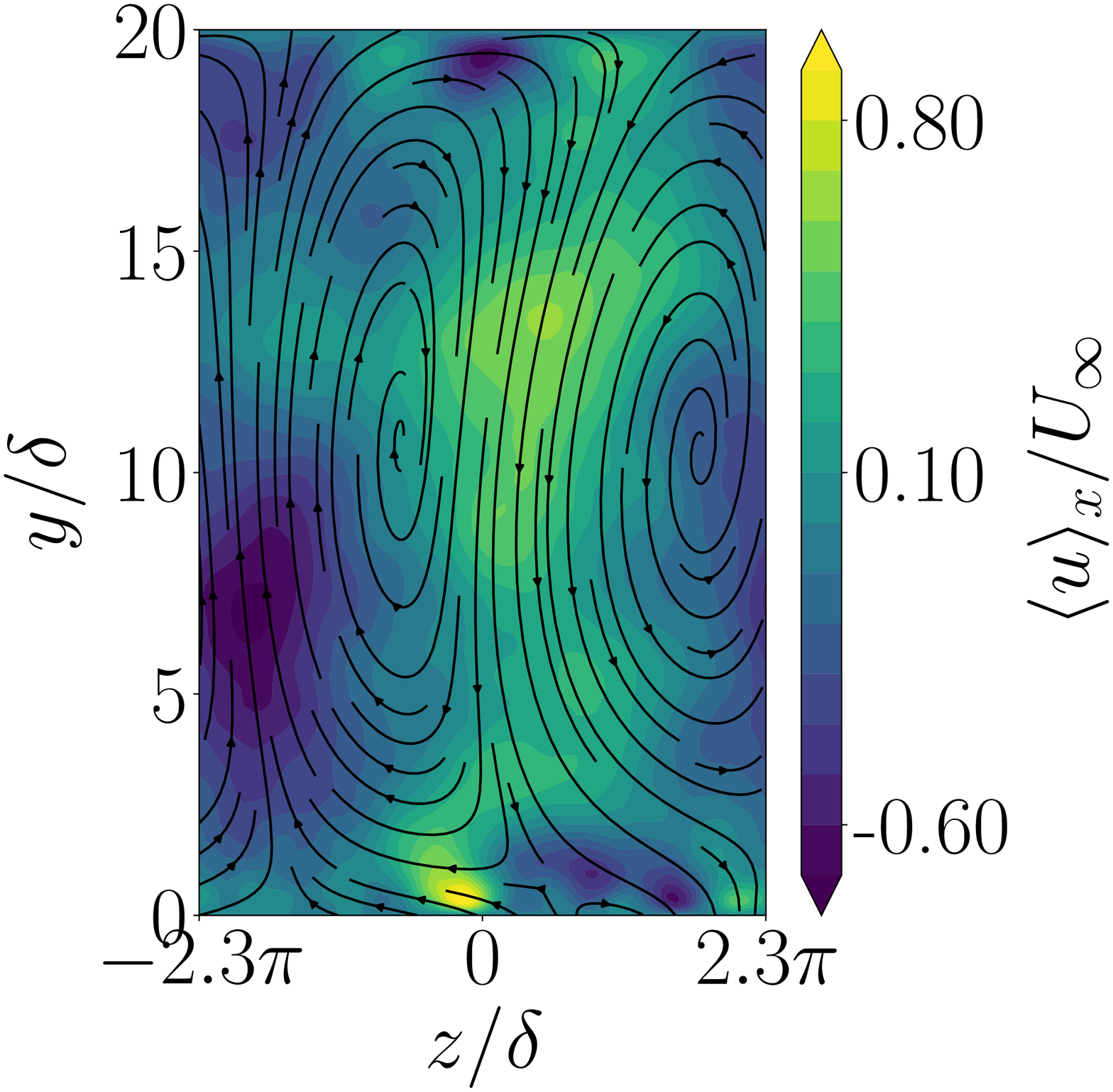}
\end{subfigure}
\end{minipage}
\begin{minipage}[b]{0.32\textwidth}
\begin{subfigure}{\textwidth}
\subcaption{}
\includegraphics[width=\linewidth]{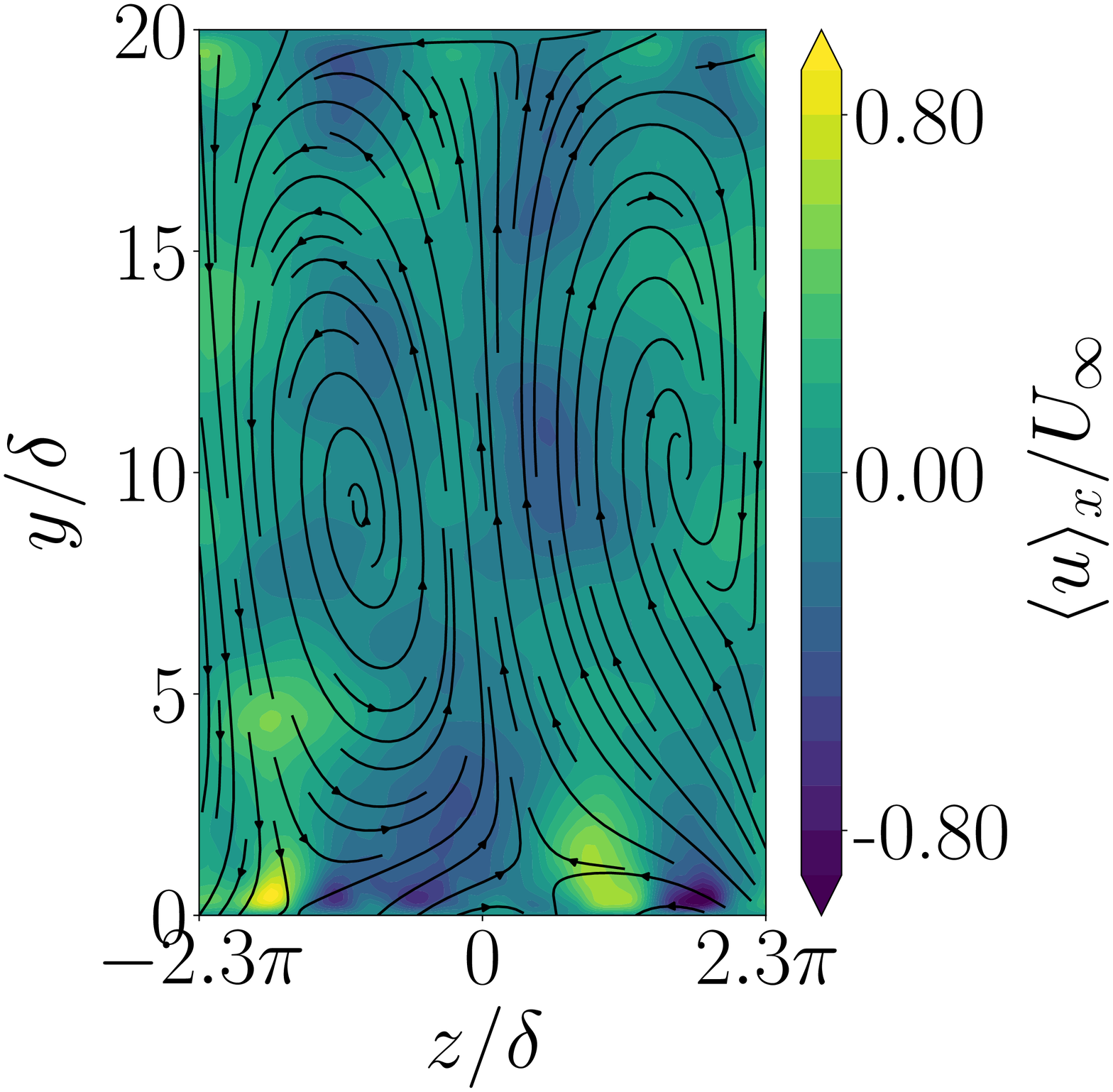}
\end{subfigure}
\end{minipage}
\begin{minipage}[b]{0.32\textwidth}
\begin{subfigure}{\textwidth}
\subcaption{}
\includegraphics[width=\linewidth]{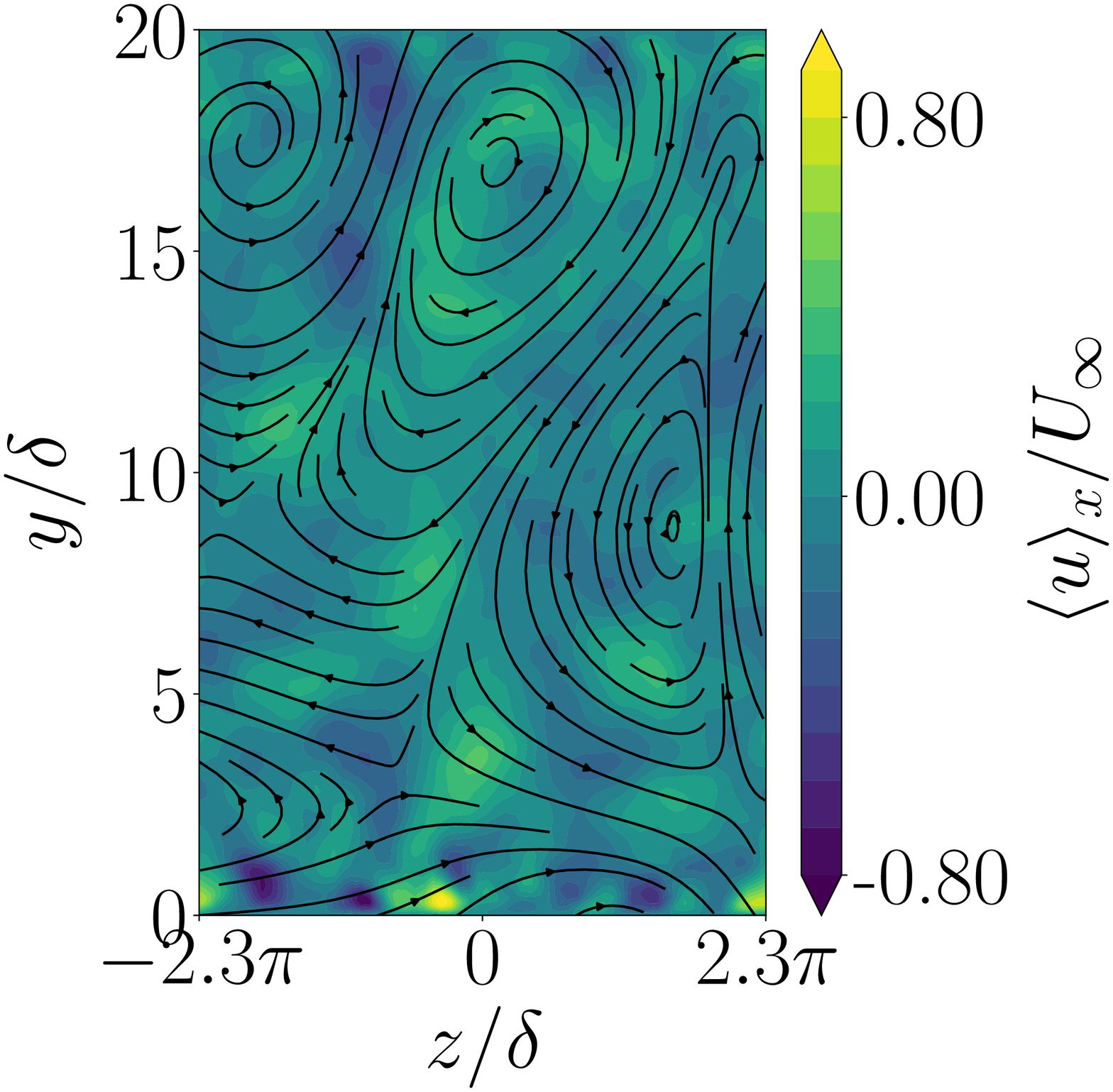}
\end{subfigure}
\end{minipage}
\quad
\begin{minipage}[b]{\textwidth}
\caption{(a) Neutral mode, (b-d) $1^{\text{th}}$-$3^{\text{th}}$ mode, (e) $11^{\text{th}}$ mode and the (f) $24^{\text{th}}$ mode with streamwise-averaged streamfunctions. The colour code indicates the streamwise-averaged deviation of the streamwise velocity from the laminar profile normalized from $-1$ to $1$.}\label{fig:spdmd-dmd-modes}
\end{minipage}
\end{figure}

In all reconstructions the sparsity-promoting DMD selects the mean flow. In the reconstruction based on seven dynamic modes the sparsity-promoting DMD optimisation selects additionally three complex conjugate pairs of dynamic modes related to the first three lowest frequencies. These frequencies consist of a fundamental frequency and its two first harmonics with a residual for the estimated fundamental frequency of $\epsilon_{\omega}\approx 2.5\cdot 10^{-3}$ using eq.~(\ref{equ:residualTg}). This property gets lost when reconstructing with thirteen or nineteen dynamic modes, where the selected frequencies are still distributed over the unit circle but are no longer commensurate and large gaps occur between adjacently selected frequencies. 

 Figure \ref{fig:spdmd-dmd-modes} illustrates a selection of dynamic modes corresponding to eigenvalues in Figure \ref{fig:eigenvalues-spdmd-asbl} as contour plots with a colour code indicating $\langle u \rangle_x$ and the streamlines the streamwise-averaged cross-flow. 
 The colour coding has been normalised between $-1$ and 1 for presentational purposes except for neutral mode, i.e. the mean flow as a time average of all snapshots over the full trajectory, shown in subfigure (a).
 This dynamic mode is clearly dominated by the large-scale low-momentum region shown in Figure~\ref{fig:3dasbl}, representing the persistence of the structure over time.
 In summary, the low-momentum region is mainly captured by the neutral mode, large-scale structures are described by the modes shown in subfigures (b) - (d), and with increasing frequency the contributions happen on smaller spatial scales with high intensities near the bottom wall, see subfigures (e) and (f). 

 The dynamic modes corresponding to the few lowest frequencies are dominated by ejection-type $(u<0,v>0)$ and sweep-type $(u>0,v<0)$ structures, as can be seen by inspection of the streamwise velocity and the cross flow in Figures~\ref{fig:spdmd-dmd-modes} (b,c) that shows the $1^{\text{st}}$ and $3^{\text{rd}}$ dynamic mode, respective. These structures are of considerable streamwise extent (not shown). 
 At high Reynolds numbers, sweep and ejection events are known to be the main contributors to the Reynolds stress distribution in turbulent boundary layers~\cite{wallace_eckelmann_brodkey_1972, Lu1973, alfredsson_johansson_1984, nolan_walsh_mceligot_2010, Jimenez2018,Lozano-Duran2012}. They are connected with turbulence production and self-contained bursts that generate streaks \cite{Jimenez2018, Lozano-Duran2012,Jimenez2022}. As such, it is conceivable that the large-scale coherent structure observed here, which indeed is a streak, is maintained by a large-scale process.  
 Figures~\ref{fig:spdmd-dmd-modes} (e) and (f) present the $11^{\text{th}}$ and $24^{\text{th}}$ dynamic mode, as can be seen these describe small wall-attached structures. 

\begin{figure}[t]
\begin{minipage}[b]{0.32\textwidth}
\begin{subfigure}{\textwidth}
\subcaption{}
\includegraphics[width=\linewidth]{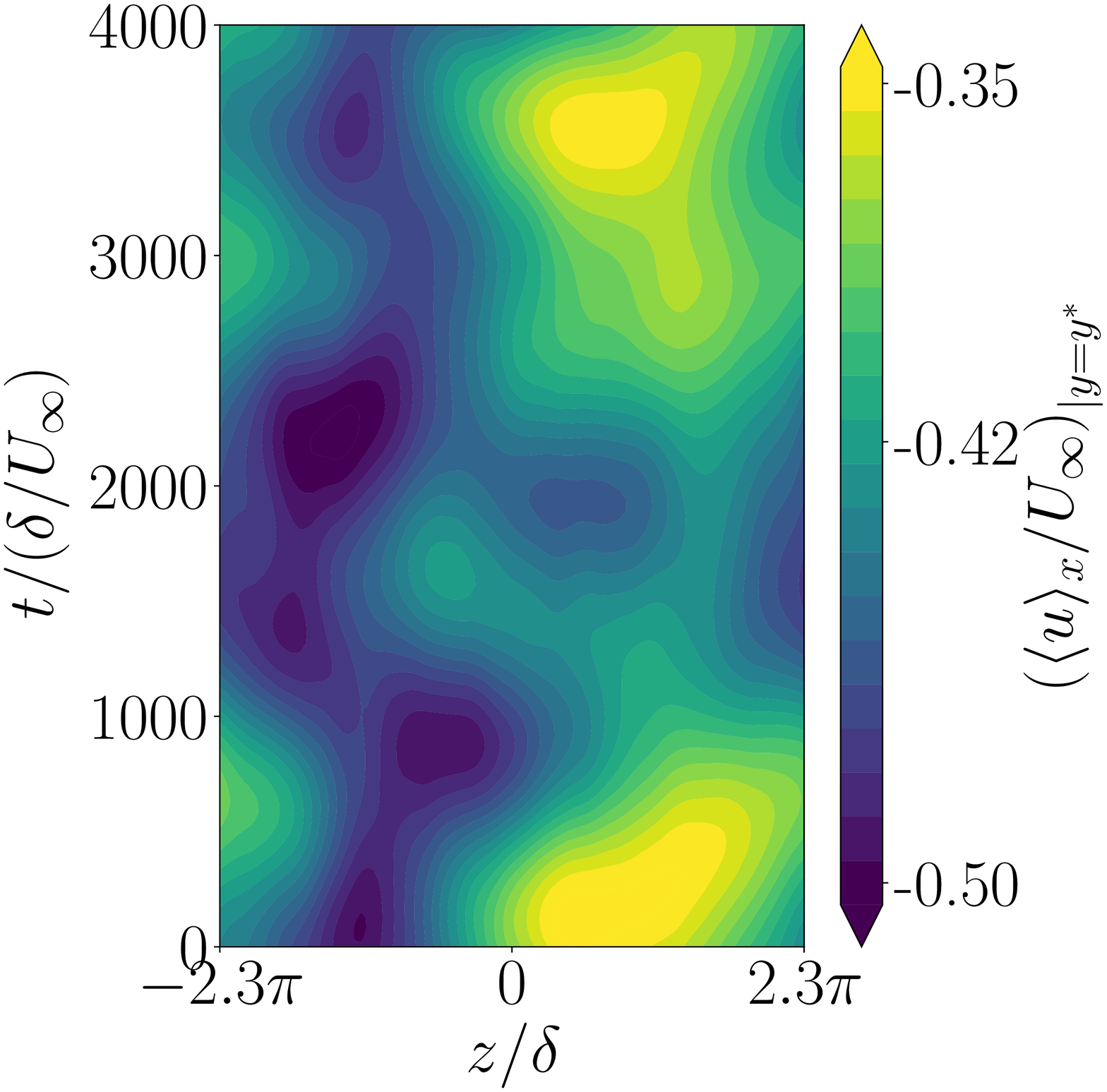}
\end{subfigure}\par
\end{minipage}
\begin{minipage}[b]{0.32\textwidth}
\begin{subfigure}{\textwidth}
\subcaption{}
\includegraphics[width=\linewidth]{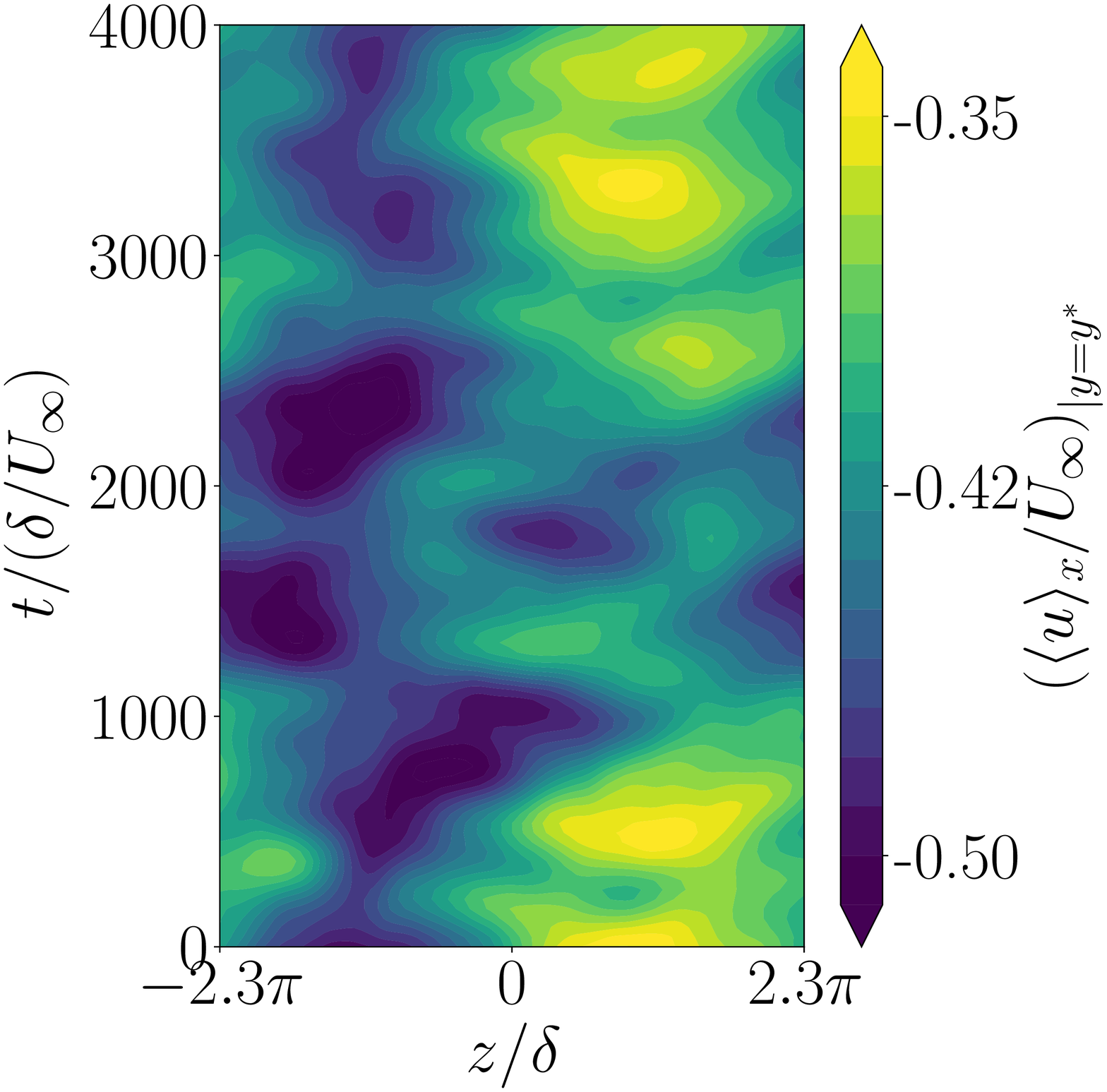}
\end{subfigure}
\end{minipage}
\begin{minipage}[b]{0.32\textwidth}
\begin{subfigure}{\textwidth}
\subcaption{}
\includegraphics[width=\linewidth]{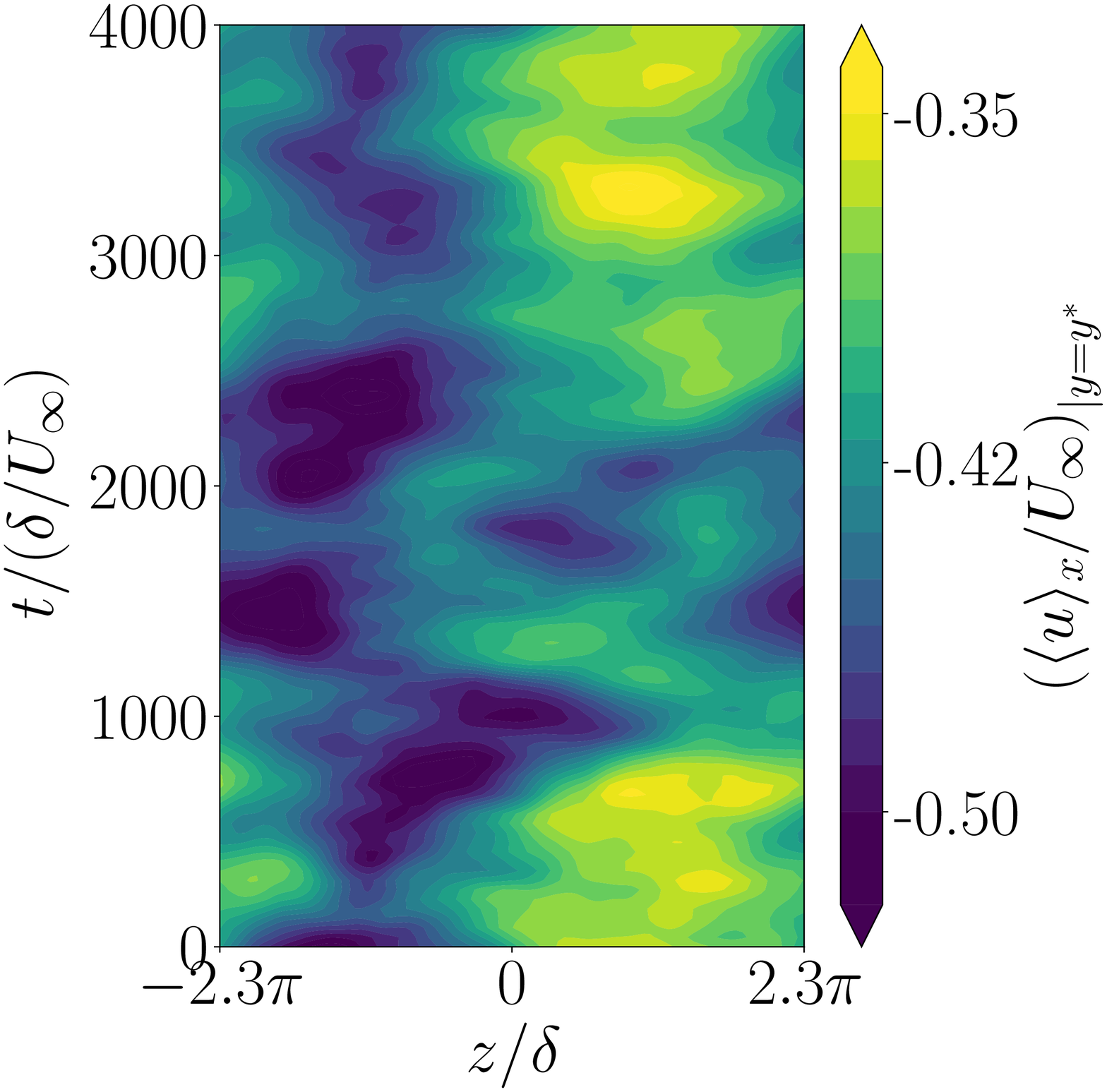}
\end{subfigure}
\end{minipage}
\quad
\begin{minipage}[b]{\textwidth}
\caption{Spatio-temporal structure of reconstructed large-scale dynamics using (a) three, (b) six and (c) nine dynamic modes. The colour code indicates the streamwise averaged deviation of the streamwise velocity from the laminar profile.}\label{fig:ucontour-symred-reconstruction}
\end{minipage}
\end{figure}

 Figure \ref{fig:ucontour-symred-reconstruction} shows the spatio-temporal structure of the reconstructed large-scale low-momentum region at wall-normal height $y/\delta\approx 3$, using seven, 13 and 19 dynamic modes corresponding to the eigenvalues shown in Figure~\ref{fig:eigenvalues-spdmd-asbl}, as a function of the spanwise coordinate and time. Again, the colour coding corresponds to $\langle u \rangle_x$. As can be seen by comparison with the space-time diagram of the symmetry-reduced data shown in Figure~\ref{fig:ucontour-nonsymred}(c), all low-dimensional representations capture the time-evolution of the large-scale low-momentum region including variations in intensity and its persistence over the entire trajectory. Furthermore, the complicated process leading to the occurrence of the secondary low-momentum region at about $t\approx 1600\delta/U_{\infty}$ and its disappearance at around $t\approx 2200\delta/U_{\infty}$ are captured in all low-dimensional representations. This is supported by the visualisations of the dynamic modes in Figure \ref{fig:spdmd-dmd-modes}. The $1^{\text{st}}$ and $2^{\text{nd}}$ dynamic modes indicate a velocity distribution, which, when evolving in time, could lead to the occurrence of the secondary low-momentum region at the spanwise coordinate $z/\delta\approx 0$. Interestingly, for the reconstruction with 19 modes the highest frequency has been selected by sparsity promotion, and a large gap between it and the next lowest frequency occurs.


\section{Conclusions}
\label{sec:conclusions}

Here, we used symmetry-reduced sparsity-promoting DMD to construct low-dimensional representations of the dynamics of a large-scale coherent structure in the ASBL and potential interactions with small-scale structures attached to the wall.  
The observed structure drifts through the simulation domain, this drift is time-dependent and reverses its direction. As such continuous symmetries are problematic for DMD-based analyses, we used symmetry-reduced DMD \cite{marensi_yalniz_hof_budanur_2023}, a recently developed method which combines DMD
with dynamical symmetry reduction using the method of slices. To obtain optimal reconstructions, we coupled SRDMD with sparsity promotion.  


We first demonstrated the symmetry-reduced sparsity-promoting DMD on a simple ASBL test case. This system with smaller computational domain and at lower Reynolds number in comparison to the system of our main interest, provided a stable relative periodic orbit consisting of a similar low-momentum region and in particular, a similar but less complicated time-dependent spanwise drift. The test case did not exhibit a streamwise drift. We performed sparsity-promoting DMD on the original data set containing the time dependent drift as well as on the symmetry-reduced data set and demonstrated the deteriorating impact the time-dependent drift has on the DMD spectrum and the low-dimensional representation. The analysis of the obtained DMD spectrum also demonstrates  the advantage symmetry reduction with the first Fourier mode slicing technique can have when DMD-based approximations are used to initialise Newton searches for relative periodic orbits~\cite{page_kerswell_2020,marensi_yalniz_hof_budanur_2023}.

We then applied sparsity-promoting SRDMD to turbulent ASBL. First, the dynamically irrelevant time-dependent transverse drift through the computational domain was removed using the first Fourier mode slicing technique with a template profile function that was a best fit to the mean velocity profile in streamwise direction. Subsequently, we constructed optimal low-dimensional approximations of the flow using seven, 13 and 19 dynamic modes using sparsity promotion. 

The reconstruction using only seven dynamic modes, related to frequencies consisting of a fundamental frequency and its first two higher harmonics captures the low-momentum region with its dominant features. This highlights the low-dimensional nature of the dynamics of this coherent structure, which does not include  any contribution from small-scale near-wall structures. It is thus conceivable that the large-scale dynamics is self-supporting. 
However, interaction between slow and the very fastest dynamics also occurs already in a low-dimensional description of the flow, here captured in the reconstructions with 13 and 19 dynamic modes. This result suggests an interplay between slow and fast dynamics at an already low-dimensional description of the flow, which remains open to study in a more detailed investigation. 

\section*{Acknowledgements}
We thank Tobias Schneider, Nazmi Burak Budanur, Elena Marensi and Jacob Page for helpful discussions and Enlin Shen for preliminary calculations. The computational resources on Cirrus ({\tt www.cirrus.ac.uk}) have been obtained through Scottish Academic Access. This work received funding from Priority Programme SPP 1881 ``Turbulent Superstructures" of the Deutsche Forschungsgemeinschaft (DFG, grant number Li3694/1). Matthias Engel was also supported by the MAC-MIGS Centre for Doctoral training at the University of Edinburgh and Heriot-Watt University through EPSRC grant EP/S023291/1. Omid Ashtari was supported by the European Research Council (ERC) under the European Union’s Horizon 2020 research and innovation programme (grant agreement no. 865677, T. Schneider).
 
\bibliographystyle{elsarticle-num} 
\bibliography{references}

\end{document}